\documentclass[twocolumn,amsmath,floatfix]{revtex4}
\usepackage{latexsym}
\usepackage{amssymb}
\usepackage{pdfpages}
\usepackage{graphics}

\begin{document}
\newcommand{\ja}{Jakuba\ss a-Amundsen }
\newcommand{\bfx}{\mbox{\boldmath $x$}}
\newcommand{\bfq}{\mbox{\boldmath $q$}}
\newcommand{\bfnabla}{\mbox{\boldmath $\nabla$}}
\newcommand{\bfalpha}{\mbox{\boldmath $\alpha$}}
\newcommand{\bfsigma}{\mbox{\boldmath $\sigma$}}
\newcommand{\bfeps}{\mbox{\boldmath $\epsilon$}}
\newcommand{\bfA}{\mbox{\boldmath $A$}}
\newcommand{\bfP}{\mbox{\boldmath $P$}}
\newcommand{\bfe}{\mbox{\boldmath $e$}}
\newcommand{\bfn}{\mbox{\boldmath $n$}}
\newcommand{\bfW}{{\mbox{\boldmath $W$}_{\!\!rad}}}
\newcommand{\bfM}{\mbox{\boldmath $M$}}
\newcommand{\bfI}{\mbox{\boldmath $I$}}
\newcommand{\bfQ}{\mbox{\boldmath $Q$}}
\newcommand{\bfp}{\mbox{\boldmath $p$}}
\newcommand{\bfk}{\mbox{\boldmath $k$}}
\newcommand{\bfks}{\mbox{{\scriptsize \boldmath $k$}}}
\newcommand{\bfqs}{\mbox{{\scriptsize \boldmath $q$}}}
\newcommand{\bfxs}{\mbox{{\scriptsize \boldmath $x$}}}
\newcommand{\bfs}{\mbox{\boldmath $s$}_0}
\newcommand{\bfv}{\mbox{\boldmath $v$}}
\newcommand{\bfw}{\mbox{\boldmath $w$}}
\newcommand{\bfb}{\mbox{\boldmath $b$}}
\newcommand{\bfxi}{\mbox{\boldmath $\xi$}}
\newcommand{\bfzeta}{\mbox{\boldmath $\zeta$}}
\newcommand{\bfzetas}{\mbox{\scriptsize \boldmath $\zeta$}}
\newcommand{\bfr}{\mbox{\boldmath $r$}}
\newcommand{\bfrs}{\mbox{{\scriptsize \boldmath $r$}}}

\renewcommand{\theequation}{\arabic{section}.\arabic{equation}}
\renewcommand{\thesection}{\arabic{section}}
\renewcommand{\thesubsection}{\arabic{section}.\arabic{subsection}}

\title{\Large\bf Radiative corrections to the spin asymmetry in elastic polarized-electron---nucleus collisions at high energies}

\author{D.~H.~Jakubassa-Amundsen \\
Mathematics Institute, University of Munich, Theresienstrasse 39,\\ 80333 Munich, Germany}

\date{\today}

\vspace{1cm}

\begin{abstract}  
Modifying the numerical codes, dispersion corrections to the beam-normal spin asymmetry which arise from low-lying transient nuclear excitations up to 30 MeV, are estimated for collision energies between 50 MeV and  1 GeV.
A nonperturbative calculation of vacuum polarization and the vertex plus self-energy correction,
using optimized potentials,
indicates that for small scattering angles both these quantum electrodynamical (QED) effects on the spin asymmetry decrease with energy above 200 MeV and can
  be neglected at high energies.
Examples are given for the $^{12}$C and $^{208}$Pb nuclei.
The available measurements of the spin asymmetry at collision energies beyond 500 MeV
cannot be explained by the present theory.
\end{abstract}

\maketitle

\section{Introduction}

Measurements of the beam-normal spin asymmetry (also known as Sherman function or vector analyzing power)
in $e + p$ collisions at 200 MeV \cite{We01} revealed a considerably larger asymmetry than predicted from elastic electron
scattering based on the phase-shift analysis \cite{CH05}.
Such large asymmetries, also found in experiments on heavier targets and collision energies above 500 MeV \cite{Ab12,Es18},
were attributed to dispersive effects, resulting from transient excitations of the target nucleus during the scattering process \cite{GH08}.
State-of-the-art calculations of the spin asymmetry due to dispersion involve the experimental forward Compton scattering cross section.
This setting results from the relation between the beam-normal spin asymmetry and the absorptive part of the two-photon exchange amplitude \cite{Ru71}
which in turn can be related by the optical theorem to the forward Compton scattering amplitude \cite{AM04}. 
Hadronic excitation energies beyond 135 MeV (the pion production threshold) are considered,
together with an empirical modelling to cover non-zero scattering angles \cite{AM04,GH08,Ko21}.
This hadronic model can explain the experimental spin-asymmetry results for light targets up to $^{90}$Zr \cite{Es20,An21,Ad22},
but it is at variance with the measurements for the $^{208}$Pb target.

Calculations allowing for larger scattering angles without additional approximations exist for proton targets, where intermediate excited hadronic states are explicitly taken into consideration \cite{PV04}.
For $^{12}$C and $^{208}$Pb targets, intermediate excited nuclear states of low angular momentum are accounted for \cite{Jaku22,JR23}, but spin asymmetries are only considered for collision energies up to 150 MeV.

In the present work the calculations of the dispersive spin asymmetry from nuclear excitations, based on  the second-order Born approximation, are extended to the GeV region.
In order to study the additional influence of the QED corections on the spin asymmetry, both vacuum polarization and vertex plus self-energy (vs) corrections are accounted  for nonperturbatively. This is done by including their respective potentials ($V_{\rm vac}$ and $V_{\rm vs}$) in the Dirac equation when solving for the electronic scattering states.
$V_{\rm vac}$ is the Uehling potential \cite{Ue35,Kl77}, while $V_{\rm vs}$ is constructed from the respective first-order Born amplitude \cite{Va00,BS19} by means of an inverse Fourier transformation \cite{Jaku24}.

The paper is organized as follows. Section 2 gives a short account of the theory and section 3 provides some numerical details.
Results for the radiative corrections in  $e + ^{12}$C and $e+^{208}$Pb collisions are given in section 4.
Due to a mistake in the earlier code for dispersion, which affects the magnetic contribution to the spin asymmetry (but not the cross section), some results from \cite{Jaku23,Jaku24} are also revised. 
Concluding remarks follow (section 5).
Atomic units ($\hbar=m_e=e=1)$ are used unless indicated otherwise.

\section{Theory}

We start by discussing dispersion and subsequently the QED effects from vacuum polarization and the vertex and self-energy corrections.
Finally the composition of all radiative effects is considered.
Infrared-divergent terms are omitted in the presentation of the transition amplitudes and cross  sections, since they have been found to cancel to all orders \cite{Ts61,YFS61}.

\subsection{Dispersion correction to elastic scattering}

Dispersion effects to elastic electron scattering arise from the fact that the nucleus is not inert during the scattering process, if the collision energy
is sufficiently high. The corresponding two-photon exchange process is in second-order Born approximation
described by the Feynman box diagram, see Fig.1.

\begin{figure}
\includegraphics[width=5cm,angle=90]{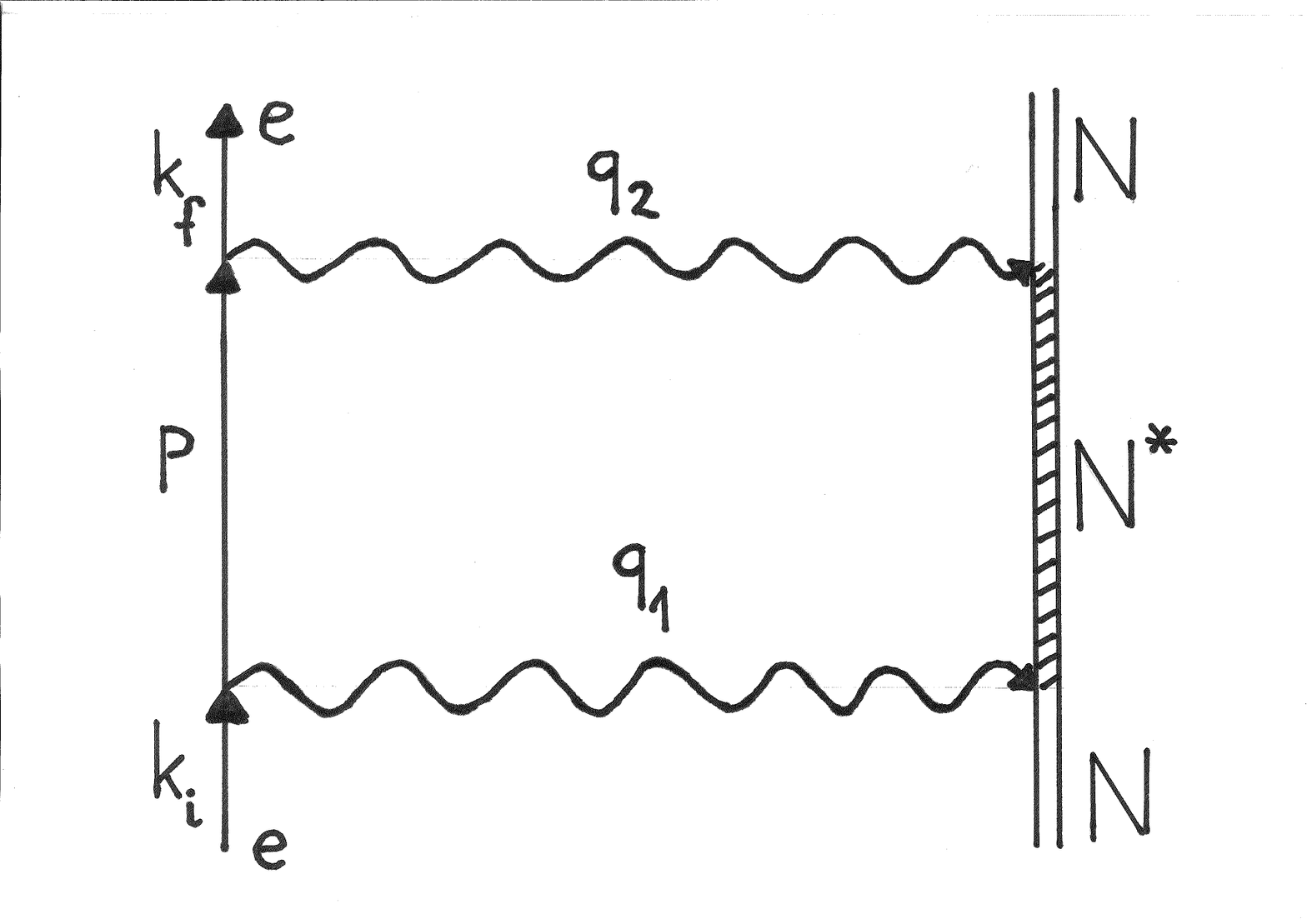}
\caption
{
Feynman box diagram. The single line represents the electron with initial, intermediate and final momenta $k_i,\;p$ and $k_f$, respectively,
and the double line represents the nucleus which is in an intermediate excited state $N^\ast$. The virtual photon momenta are denoted by $q_1$ and $q_2$. 
}
\end{figure}

The amplitude $A_{fi}^{\rm box}$ for this process can be written in the following form \cite{FR74,BD64},
$$A_{fi}^{\rm box}\,=\,\frac{\sqrt{E_iE_f}}{\pi^2c^3}
\sum_{L,\omega_L}\sum_{M=-L}^L \int d\bfp$$
\begin{equation}\label{2.1}
\times\;\sum_{\mu,\nu=0}^3 \frac{1}{(q_2^2+i\epsilon)(q_1^2+i\epsilon)}\;t_{\mu\nu}(p)\;T^{\mu\nu}(LM,\omega_L),
\end{equation}
where $E_i,k_i$ and $E_f,k_f$ are, respectively, the initial and final total energies and four-momenta of the scattering electron.
The denominator results from the propagators of the first and second photon with four-momentum $q_1=k_i-p$ and $q_2=p-k_f$, respectively.
Here,  $p=(E_p/c,\bfp)$ is the momentum of the intermediate electronic state.
The electronic transition matrix element is denoted by $t_{\mu,\nu}(p)$, while $T^{\mu\nu}(LM,\omega_L)$
represents the nuclear transition matrix element for the excitation to a state with energy $\omega_L$,
angular momentum $L$, magnetic projection $M$,
and its subsequent decay to the ground state.
Explicit formulae can be found in \cite{FR74} (where, however, an additional closure approximation is introduced) and in \cite{Jaku22}.
In principle the sum runs over all nuclear excited states, but can be confined to low angular momentum states if the momentum transfer
$\bfq=\bfk_i-\bfk_f$ to the nucleus is not too high.

Let us denote the exact transition amplitude for potential scattering from the nuclear Coulombic field $V_T$ by $f_{\rm coul}$. Then the differential cross section for the elastic scattering of electrons 
with spin-polarization vector $\bfzeta_i$ into the solid angle $d\Omega_f$, including dispersion to lowest order, 
is given by 
\begin{equation}\label{2.2}
\frac{d\sigma_{\rm box}}{d\Omega_f}(\bfzeta_i) \,=\,\frac{|\bfk_f|}{|\bfk_i|} \,\frac{1}{f_{\rm rec}} \sum_{\sigma_f}\left[ \,|f_{\rm coul}|^2\,+\,2\mbox{ Re } \{ f_{\rm coul}^\ast \,A_{fi}^{\rm box}\}\right], 
\end{equation}
where the leading-order term defines the Coulombic cross section $d\sigma_{\rm coul}(\bfzeta_i)/d\Omega_f$.
A sum over the final spin polarization $\sigma_f$ of the electron is included.
Recoil is considered by the prefactor $f_{\rm rec}^{-1}$ as well as by using an averaged collision energy $\sqrt{(E_i-c^2)(E_f-c^2)}$ when calculating the scattering amplitude from the phase-shift analysis \cite{Jaku22}.

The cross section change by dispersion is obtained from
\begin{equation}\label{2.3}
\Delta \sigma_{\rm box}\;=\;\frac{d\sigma_{\rm box}/d\Omega_f\,-\,d\sigma_{\rm coul}/d\Omega_f}{d\sigma_{\rm coul}/d\Omega_f},
\end{equation}
where in each cross section on the rhs of (\ref{2.3}) an average over the initial electronic spin polarization is implied.
Here and in the following it is assumed that the nuclear ground state has spin zero, such that no leading-order magnetic scattering occurs.

From the linearity of (\ref{2.2}) in $A_{fi}^{\rm box}$ it follows that $\Delta\sigma_{\rm box}$ is additive with respect to the contributing intermediate nuclear states,
\begin{equation}\label{2.4}
\Delta \sigma_{\rm box}\;=\;\sum_{L,\omega_L} \Delta \sigma_{\rm box}(L,\omega_L),
\end{equation}
where $\Delta\sigma_{\rm box}(L,\omega_L)$ denotes the modification of the cross section by considering the excitation of a single state characterized by $L$ and $\omega_L$.
 
The Sherman function $S$ for electrons polarized perpendicular to the scattering plane (such that $\bfzeta_i$ is aligned with $\bfk_i \times \bfk_f$) is defined as the relative cross-section difference
when the beam polarization is flipped \cite{Mo64},
\begin{equation}\label{2.5}
S\;=\;\frac{d\sigma/d\Omega_f(\bfzeta_i)-d\sigma/d\Omega_f(-\bfzeta_i)}{d\sigma/d\Omega_f(\bfzeta_i) + d\sigma/d\Omega_f(-\bfzeta_i)}.
\end{equation}
The spin asymmetry refers to $S_{\rm box}$ and to $S_{\rm coul}$ if, respectively, $d\sigma/d\Omega_f$ is replaced by 
$d\sigma_{\rm box}/d\Omega_f$  and $d\sigma_{\rm coul}/d\Omega_f$.
Let us denote by $S_{\rm box}(L,\omega_L)$ the spin asymmetry when only one excited state, characterized by $L$ and $\omega_L$, is considered in the cross section. Then $S_{\rm box}$ can be written in the following way,
$$
S_{\rm box}\,=\, \frac{S_{\rm coul}}{1+\Delta\sigma_{\rm box}}$$
\begin{equation}\label{2.6}
+\,\frac{\sum_{L,\omega_L} \left[ S_{\rm box}(L,\omega_L)(1+\Delta\sigma_{\rm box}(L,\omega_L))-S_{\rm coul}\right]}{1+\Delta \sigma_{\rm box}}.\end{equation}
If the dispersive cross section change is small, i.e.
$|\Delta\sigma_{\rm box}|\ll 1$ and $|\Delta\sigma_{\rm box}(L,\omega_L)|\ll 1$ for all $L,\omega_L$, the Sherman-function change is additive as well,
\begin{equation}\label{2.7}
S_{\rm box} -S_{\rm coul}\,\approx\, \sum_{L,\omega_L} \left[ S_{\rm box}(L,\omega_L)-S_{\rm coul}\right].
\end{equation}
By means of (\ref{2.7}) the absolute modification of the Sherman function is considered, as opposed to its change $dS_{\rm box}$ relative to $S_{\rm coul}$,
\begin{equation}\label{2.7a}
dS_{\rm box}\;=\;\frac{S_{\rm box} -S_{\rm coul}}{S_{\rm coul}}.
\end{equation}
Using $S_{\rm box} -S_{\rm coul}$ has the advantage that it is well defined near the zeros of $S_{\rm coul}$, and
that it can directly be compared to the Born prediction for the dispersive spin asymmetry.

\subsection{Quantum electrodynamical (QED) corrections}

Vacuum polarization and the vertex plus self-energy (vs) correction are considered nonperturbatively in terms of their respective potentials,
$V_{\rm vac}$ and $V_{\rm vs}$.  The Uehling potential for a spherical nuclear charge distribution $\varrho_N(r)$ (normalized to the nuclear charge number $Z$) is given by \cite{FR76,Kl77}
$$
V_{\rm vac}(r)\,=\, -\,\frac{2}{3c^2r} \int_0^\infty r'dr'\,\varrho_N(r')\left[ \chi_2(2c|r-r'|)
\right.$$
\begin{equation}\label{2.8}
\left.-\,\chi_2(2c|r+r'|)\right],
\end{equation}
$$\chi_n(x)\,=\,\int_1^\infty dt\,e^{-xt}\,t^{-n}\left( 1\,+\,\frac{1}{2t^2}\right)\left( 1\,-\,\frac{1}{t^2}\right)^\frac12,
$$
$n=1,2,...,$
which for large distances (compared to the radius $R_N$ of the nuclear charge distribution)
can be approximated by the point-nucleus Uehling potential \cite{FR76},
\begin{equation}\label{2.9}
V_{\rm vac}(r)\,=\,-\;\frac{2Z}{3c\pi r}\;\chi_1(2cr),\qquad r \gg R_N.
\end{equation}

The potential for the vs correction is derived from the dominant part of its first-order transition  amplitude, which is
given in terms of the electric form factor $F_1^{\rm vs}$ multiplying the first-order Born approximation $A_{fi}^{B1}$ for potential scattering \cite{MT00},	
$$
A_{fi}^{\rm vs(1)}\;=\;  F_1^{\rm vs}(-q^2)  \;A_{fi}^{B1}.
$$
$$F_1^{\rm vs}(-q^2)\;=\;\frac{1}{2\pi c}\left\{ \frac{v^2+1}{4v}\left(\ln \,\frac{v+1}{v-1}\right)\left( \ln\,\frac{v^2-1}{4v^2}\right)\right.$$
\begin{equation}\label{2.10}
+\;\frac{2v^2+1}{2v} \,\ln\,\frac{v+1}{v-1}
 -\,2\,
\end{equation}
$$\left.+\;\frac{v^2+1}{2v}\left[\mbox{Li }\left(\frac{v+1}{2v}\right)
  \;-\;\mbox{Li }\left(\frac{v-1}{2v}\right)\right]\right\},
$$
where 
Li$(x)=-\int_0^x dt \frac{\ln|1-t|}{t}\;$ is the Spence function \cite{Ts61,Va00} and
 $v=\sqrt{1-4c^2/q^2}$ with
 $q^2=(E_i-E_f)^2/c^2-\bfq^2$.
 
The  inverse Fourier transform of the product of the electric form factor $F_1^{\rm vs}$ and the nuclear charge form factor $F_L$ 
leads to the potential $V_{\rm vs}$ \cite{Jaku24},
$$
V_{\rm vs}(r)\,=\,-\,\frac{2Z}{\pi} \int_0^\infty d|\bfq|\;\frac{\sin(|\bfq|r)}{|\bfq|\,r}\;F_L(|\bfq|)\;F_1^{\rm vs}(-q^2),$$
\begin{equation}\label{2.11}
F_L(|\bfq|)\,=\,-\,\frac{\bfq^2}{4\pi Z}\int d\bfr\;e^{i\bfqs \cdot \bfrs} \,V_T(r).
\end{equation}

This allows for the calculation of the electronic scattering state $\psi(\bfr)$ with total energy $E$
from the Dirac equation under the influence of the combined potentials,
\begin{equation}\label{2.12}
\left[ -ic\bfalpha \bfnabla + \gamma_0 c^2 + V_T(r) +V_{\rm vac}(r)+V_{\rm vs}(r) \right]\psi(\bfr)=E\,\psi(\bfr),
\end{equation}
where $\bfalpha$ and $\gamma_0$ refer to  Dirac matrices \cite{BD64}.
Applying  the phase-shift analysis to $\psi$ as done for the Coulombic scattering leads to the transition amplitude $f_{\rm vac+vs}$,
which includes the vacuum polarization and the vs correction nonperturbatively.

There is an additional (magnetic) contribution $A_{fi}^{\rm vs(2)}$ to the vs transition amplitude \cite{BS19} which is usually negligible (except for very small momentum transfer),
$$A_{fi}^{\rm vs(2)}\,=\,\frac{1}{2c}\;F_2^{\rm vs}(-q^2)\;\frac{(u_{k_f}^{+(\sigma_f)}\gamma_0 (\bfalpha \bfq)\,u_{k_i}^{(\sigma_i)})}{(u_{k_f}^{+(\sigma_f)}\,u_{k_i}^{(\sigma_i)})}\;f_{\rm coul},$$
\begin{equation}\label{2.13}
F_2^{\rm vs}(-q^2)\;=\;\frac{1}{4\pi c}\;\frac{v^2-1}{v}\;\ln \frac{v+1}{v-1}.
\end{equation}
Since $A_{fi}^{\rm vs(2)}$ does not allow to define a suitable potential, it has to be included perturbatively. Therefore
 the  Born amplitude $A_{fi}^{B1}$ in the original definition of $A_{fi}^{\rm vs(2)}$ is replaced in (\ref{2.13}) by $f_{\rm coul}$ in order to account for the Coulomb distortion (as suggested in \cite{Ma69}),
and $u_k^{(\sigma)}$ denotes the free four-spinor for an electron with momentum $k$ and spin polarization $\sigma$ \cite{BD64}.

Soft bremsstrahlung is considered by means of its cross section \cite{MT00}
$$ \frac{d\sigma^{\rm soft}}{d\Omega_f}\;=\;W_{fi}^{\rm soft}\;\left| f_{\rm vac+vs}\right|^2,$$
\begin{equation}\label{2.14}
W_{fi}^{\rm soft}\,=\,\frac{1}{\pi c} \left\{ [\ln(-q^2/c^2)-1]\;\ln\frac{\omega_0^2}{E_iE_f}\,+\,\frac12\left( \ln(-q^2/c^2)\right)^2\right.
\end{equation}
$$\left. -\,\frac12\;\left( \ln\frac{E_i}{E_f}\right)^2\,+\,\mbox{Li} (\cos^2(\vartheta_f/2))\,-\frac{\pi^2}{3}\right\},$$
valid for sufficiently large momentum transfer ($-q^2/c^2 \gg 1$).
The scattering angle is denoted by $\vartheta_f$, and $\omega_0$, characterizing the cut-off frequency for the soft bremsstrahlung,
is given by the resolution $\Delta E$ of the electron detector.
The factor $|f_{\rm vac+vs}|^2$ in place of $|A_{fi}^{B1}|^2$ from the Born approximation accounts for the fact that the additional photon emission in a given scattering process (which is assumed to take place in the combined potential
$V_T+V_{\rm vac}+V_{\rm vs}$) is represented by the cross section for this scattering process, multiplied by a factor which describes the attachment of an extra photon line to the respective diagram \cite{We65}.

Omitting $A_{fi}^{\rm vs(2)} $, the differential cross section for elastic electron scattering in the presence of these QED effects reads
\begin{equation}\label{2.15}
\frac{d\sigma_{\rm QED}}{d\Omega_f}(\bfzeta_i)\,\approx\, \left( 1\,+\,W_{fi}^{\rm soft}\right)\,\frac{|\bfk_f|}{|\bfk_i|}\;\frac{1}{f_{\rm rec}}\sum_{\sigma_f} \left| f_{\rm vac+vs}\right|^2,
\end{equation}
where it is profitted from the spin-independence of $W_{fi}^{\rm soft}$.
For energies well below 20 MeV, the magnetic contribution $A_{fi}^{\rm vs(2)}$ can no longer be neglected. Including it to lowest order, one obtains
$$
\frac{d\sigma_{\rm QED}}{d\Omega_f}(\bfzeta_i) \,=\,\frac{|\bfk_f|}{|\bfk_i|} \,\frac{1}{f_{\rm rec}}(1+W_{fi}^{\rm soft}) \sum_{\sigma_f}\left[ \,|f_{\rm vac+vs}|^2
\right.$$
\begin{equation}\label{2.16}
\left. +\;2\mbox{ Re }\{f_{\rm coul}^\ast A_{fi}^{\rm vs(2)}\} \right].
\end{equation}
For being consistent to lowest order, the term proportional to $W_{fi}^{\rm soft}\cdot A_{fi}^{\rm vs(2)}$ can be dropped when calculating the cross section. However, it has to be kept when calculating the spin asymmetry.

The cross section change by the QED effects is obtained from
\begin{equation}\label{2.17}
\Delta\sigma_{\rm QED}\;=\;\frac{d\sigma_{\rm QED}/d\Omega_f-d\sigma_{\rm coul}/d\Omega_f}{d\sigma_{\rm coul}/d\Omega_f},
\end{equation}
where again an average over the initial electronic spin  polarization is implemented.

The Sherman function is obtained by the formula (\ref{2.5}),
from which it follows that $(1+W_{fi}^{\rm soft}$) drops out when calculating the spin asymmetry. 
When $A_{fi}^{\rm vs(2)}$ is negligible, $S_{\rm QED}$ can alternatively be obtained from the formula
provided by the phase-shift analysis,
\begin{equation}\label{2.18}
S_{\rm QED} \,\approx\, \frac{2\mbox{ Re } AB^\ast}{|A|^2+|B|^2},
\end{equation}
where $A$ is the direct and $B$ the spin-flip part of the scattering amplitude $f_{\rm vac+vs}$
 \cite{Mo64,Lan}.
In particular, $S_{\rm QED}$ is independent of bremsstrahlung and of the detector resolution as already noted in \cite{Jo62}.

\subsection{Combined radiative corrections}

The differential cross section for elastic scattering, including dispersion and the QED corrections, is given by the sum of the respective contributions,
$$\frac{d\sigma_{\rm tot}}{d\Omega_f}(\bfzeta_i)\;=\;\frac{d\sigma_{\rm QED}}{d\Omega_f}(\bfzeta_i)\,+\,\frac{D\sigma_{\rm box}}{d\Omega_f}(\bfzeta_i)$$
\begin{equation}\label{2.19}
\frac{D\sigma_{\rm box}}{d\Omega_f}(\bfzeta_i)\;=\;\frac{|\bfk_f|}{|\bfk_i|}\;\frac{1}{f_{\rm rec}}\;(1+W_{fi}^{\rm soft}) \sum_{\sigma_f}2\mbox{ Re }\{f^\ast_{\rm coul}A_{fi}^{\rm box}\},
\end{equation}
with $d\sigma_{\rm QED}/d\Omega_f$ from (\ref{2.16}).
The Sherman function is
calculated from
\begin{equation}\label{2.20}
S_{\rm tot}\,=\,\frac{\frac{d\sigma_{\rm QED}}{d\Omega_f}(\bfzetas_i)+\frac{D\sigma_{\rm box}}{d\Omega_f}(\bfzetas_i) -\frac{d\sigma_{\rm QED}}{d\Omega_f}(-\bfzetas_i)-\frac{D\sigma_{\rm box}}{d\Omega_f}(-\bfzetas_i)}
{\frac{d\sigma_{\rm QED}}{d\Omega_f}(\bfzetas_i)+\frac{D\sigma_{\rm box}}{d\Omega_f}(\bfzetas_i)+\frac{d\sigma_{\rm QED}}{d\Omega_f}(-\bfzetas_i)+\frac{D\sigma_{\rm box}}{d\Omega_f}(-\bfzetas_i)}.
\end{equation}
Thus the factor $(1+W_{fi}^{\rm soft})$ drops out and the Sherman function remains independent of bremsstrahlung.

 For simplifying the notation we abbreviate in the following $\frac{d\sigma}{d\Omega}(\bfzetas_i)\pm \frac{d\sigma}{d\Omega}(-\bfzetas_i)$ by $\frac{d\sigma}{d\Omega}(\uparrow \pm \downarrow)$.

When the cross section change by dispersion is small ($\frac{D\sigma_{\rm box}}{d\Omega_f}\ll \frac{d\sigma_{\rm coul}}{d\Omega_f},\;\frac{D\sigma_{\rm box}}{d\Omega_f} \ll \frac{d\sigma_{\rm QED}}{d\Omega_f}$),
the denominator of (\ref{2.20}) can be expanded to first order. Using $S_{\rm QED}=\frac{d\sigma_{\rm QED}/d\Omega_f(\uparrow - \downarrow)}{d\sigma_{\rm QED}/d\Omega_f(\uparrow +\downarrow)}$, one obtains
$$S_{\rm tot} \,\approx\, S_{\rm QED}\left(1\,-\,\frac{D\sigma_{\rm box}/d\Omega_f(\uparrow +\downarrow)}{d\sigma_{\rm QED}/d\Omega_f(\uparrow +\downarrow)}\right)$$
\begin{equation}\label{2.21}
+\;\frac{D\sigma_{\rm box}/d\Omega_f(\uparrow - \downarrow)}{d\sigma_{\rm QED}/d\Omega_f(\uparrow +\downarrow)}.
\end{equation}
In the following equations the factor $(1+W_{fi}^{\rm soft})$ is replaced by unity in (\ref{2.16}) and (\ref{2.19}) for the calculation of the Sherman function. Then one has
\begin{equation}\label{2.21a}
\frac{D\sigma_{\rm box}}{d\Omega_f}(\bfzeta_i)\;=\;\frac{d\sigma_{\rm box}}{d\Omega_f}(\bfzeta_i)\,-\,\frac{d\sigma_{\rm coul}}{d\Omega_f}(\bfzeta_i).
\end{equation}
Therefore $\frac{D\sigma_{\rm box}}{d\Omega_f}(\uparrow - \downarrow)$ can be  expressed with the help of $S_{\rm box}$ by means of
\begin{equation}\label{2.22}
\frac{\frac{D\sigma_{\rm box}}{d\Omega_f}(\uparrow - \downarrow)}{\frac{d\sigma_{\rm box}}{d\Omega_f}(\uparrow +\downarrow)}\,=\,S_{\rm box}\,-\,S_{\rm coul}\frac{\frac{d\sigma_{\rm coul}}{d\Omega_f}(\uparrow+\downarrow)}{\frac{d\sigma_{\rm box}}{d\Omega_f}(\uparrow +\downarrow)}.
\end{equation}
This leads to
$$S_{\rm tot}\,-\,S_{\rm coul}\;\approx$$
\begin{equation}\label{2.23}
S_{\rm QED}\left( 1\,-\,\frac{\frac{d\sigma_{\rm box}}{d\Omega_f}(\uparrow + \downarrow)-\frac{d\sigma_{\rm coul}}{d\Omega_f}(\uparrow + \downarrow)}{\frac{d\sigma_{\rm QED}}{d\Omega_f}(\uparrow + \downarrow)}\right) -\;S_{\rm coul}
\end{equation}
$$+\left[S_{\rm box} -\,S_{\rm coul}\frac{\frac{d\sigma_{\rm coul}}{d\Omega_f}(\uparrow+\downarrow)}{\frac{d\sigma_{\rm box}}{d\Omega_f}(\uparrow + \downarrow)}\right]\;\frac{\frac{d\sigma_{\rm box}}{d\Omega_f}(\uparrow + \downarrow)}{\frac{d\sigma_{\rm QED}}{d\Omega_f}(\uparrow + \downarrow)}.$$
If one disregards the difference between $d
\sigma_{\rm coul}/d\Omega_f$ and $d\sigma_{\rm box}/d\Omega_f$
and drops the first-order expansion term of the denominator,  the formula given previously is recovered \cite{Jaku24},
$$S_{\rm tot}\,-\,S_{\rm coul}\,\approx \,S_{\rm QED}\,-\,S_{\rm coul}$$
\begin{equation}\label{2.24}
+\;\left[ S_{\rm box} -\,S_{\rm coul}\right]\;\frac{d\sigma_{\rm coul}/d\Omega_f(\uparrow + \downarrow)}{d\sigma_{\rm QED}/d\Omega_f(\uparrow + \downarrow)}.
\end{equation}

Excluding the vicinity of zeros in $S_{\rm coul}$, the relative change $dS_{\rm tot}$ of the Sherman function can be obtained
upon dividing (\ref{2.24}) by $S_{\rm coul}$. Thus one arrives at the total relative change of the spin asymmetry
by the radiative corrections,
\begin{equation}\label{2.26}
dS_{\rm tot} \;\approx\; dS_{\rm QED} \,+\,dS_{\rm box}\cdot \frac{1}{1+\Delta \sigma_{\rm QED}},
\end{equation}
with $\Delta \sigma_{\rm QED}$ from (\ref{2.17}).
Recall that  the bremsstrahlung contribution in $\Delta\sigma_{\rm QED}$  has to be omitted.
(Item (v) in \cite{Jaku24}, concerning the inaccuracy due to the  detector resolution, should be removed.)

\begin{figure}
\vspace{-1.5cm}
\includegraphics[width=11cm]{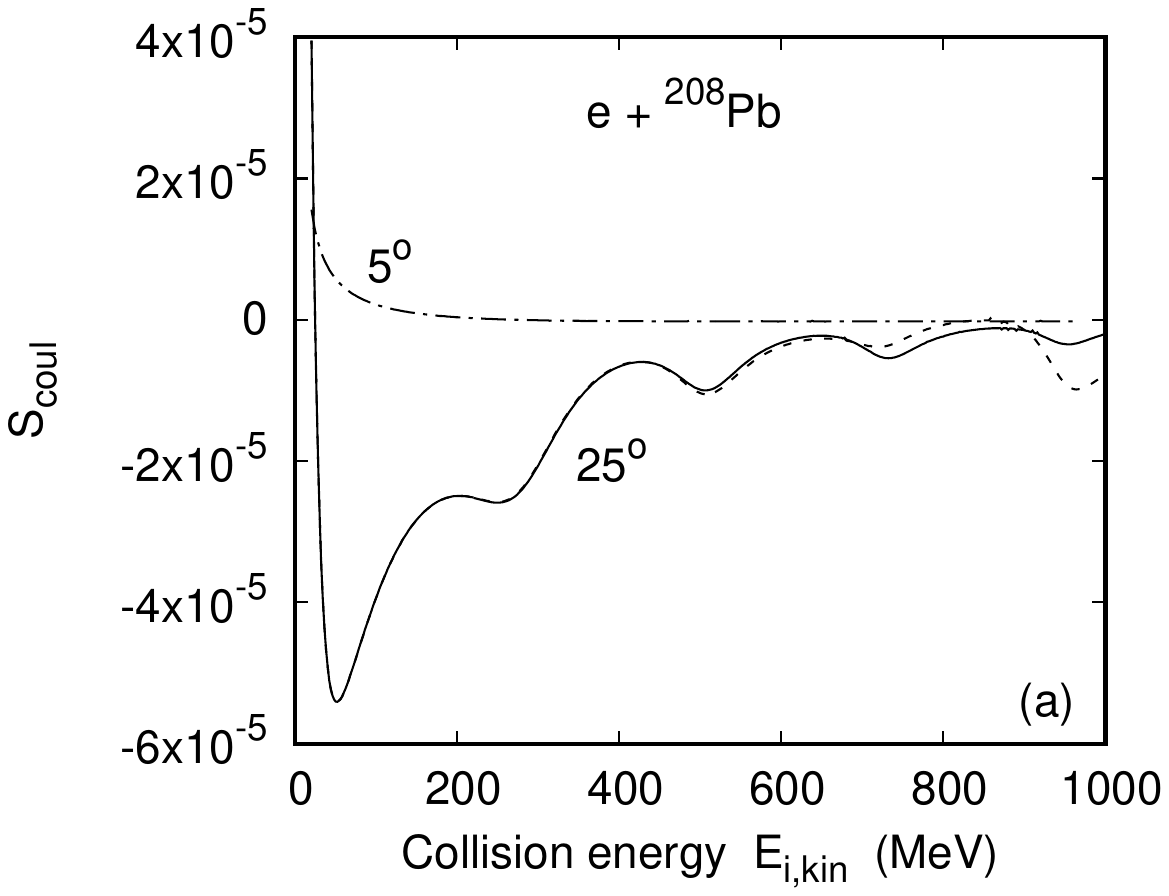}
\vspace{-1.5cm}
\vspace{-0.5cm}
\includegraphics[width=11cm]{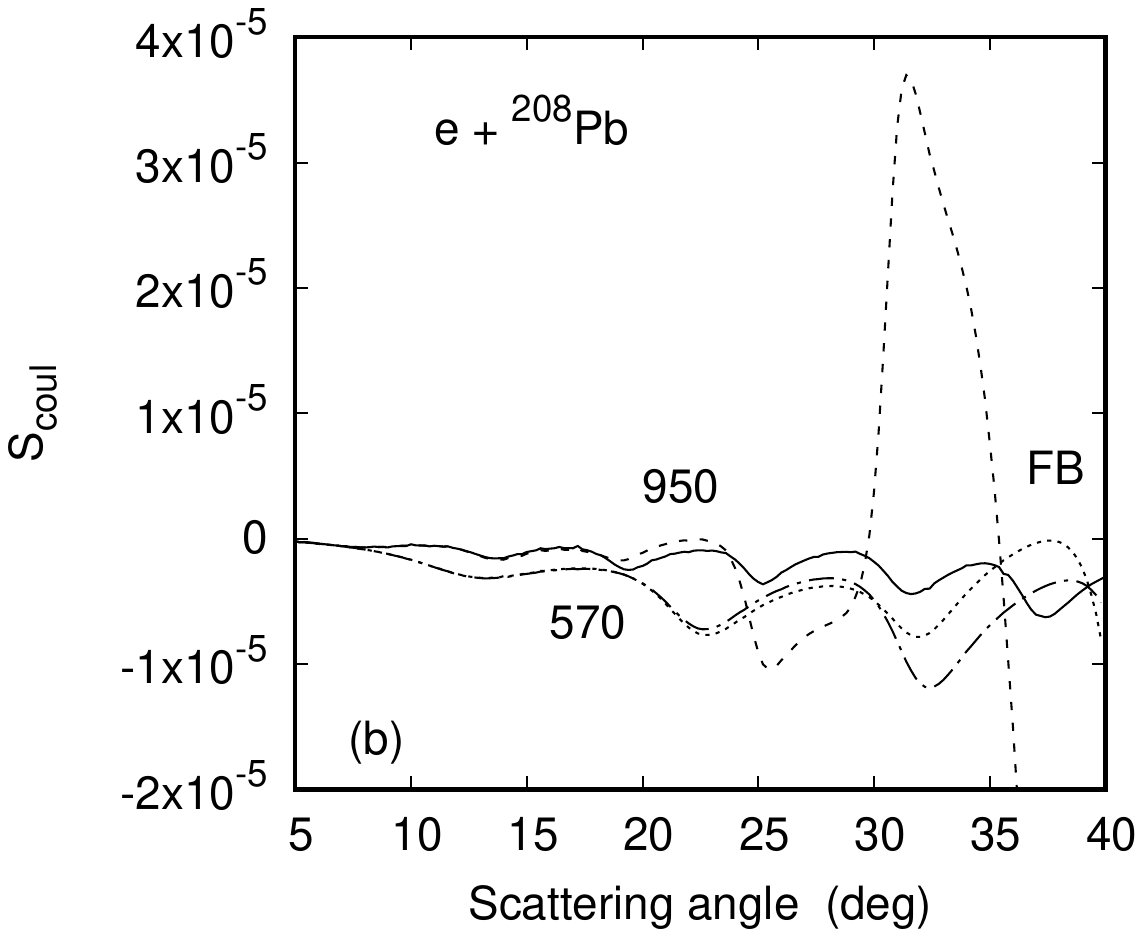}
\vspace{-1.5cm}
\vspace{-0.5cm}
\includegraphics[width=11cm]{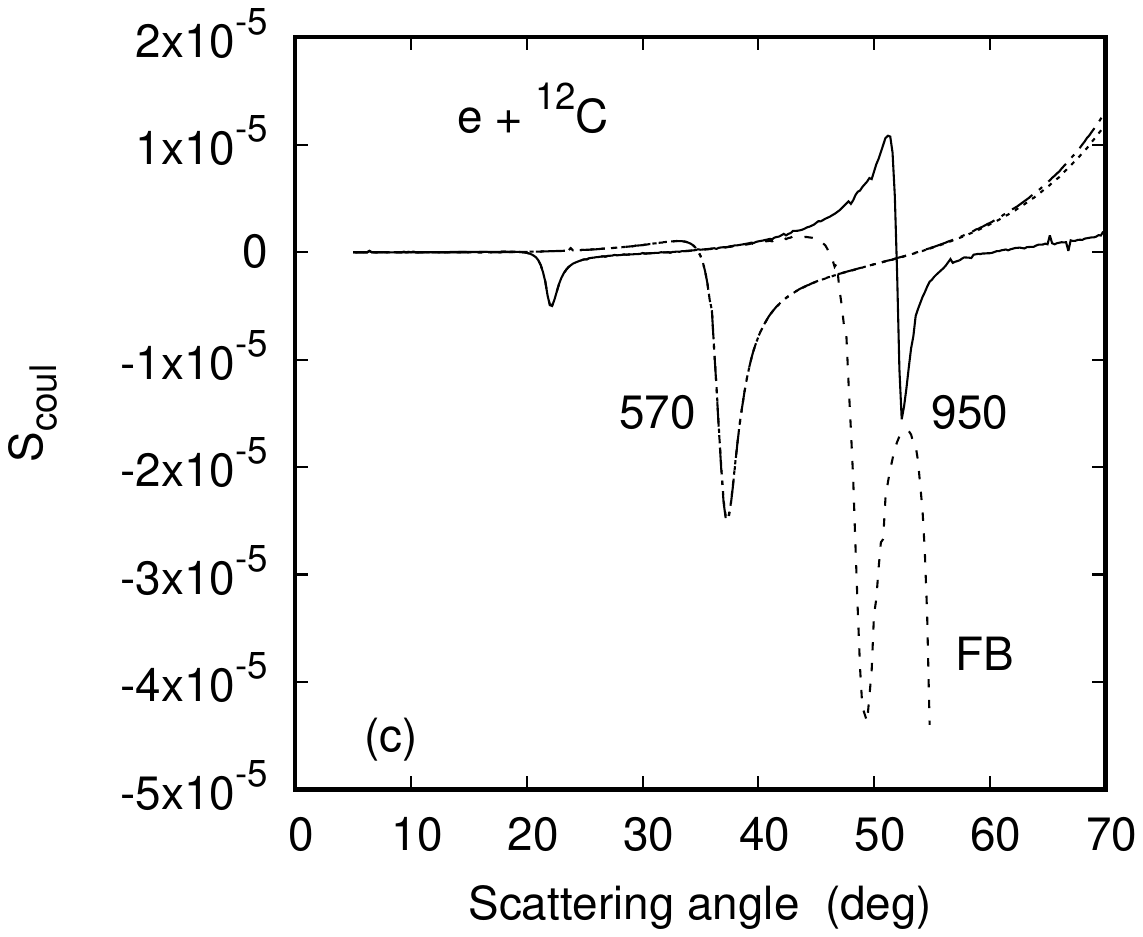}
\caption
{
Sherman function $S_{\rm coul}$ (a) in $e+^{208}$Pb collisions at $\vartheta_f =5^\circ$ and $25^\circ$ as a function of collision energy $E_{\rm i,kin}=E_i-c^2$ and (b)
in $e+^{208}$Pb and (c) in $e+^{12}$C collisions at 570 and 950 MeV as a function of scattering angle $\vartheta_f.$
Results from Gaussian charge distribution in (a) at $25^\circ$ (-------) and $5^\circ\; (-\cdot -\cdot -)$
and in (b) and (c) at 950 MeV (-------) and 570 MeV $(-\cdot -\cdot -)$.
Results from Fourier-Bessel charge distribution in (a) at $25^\circ \;(----)$ and in (b) and (c) at 950 MeV $(----)$ and 570 MeV $(\cdots\cdots)$.
}
\end{figure}

\section{Numerical details}
\setcounter{equation}{0}

In order to cover energies up to 1 GeV some improvements of the numerical QED code are necessary.
Modifications of the nuclear ground-state density, of the QED potentials and of the way of calculating potential scattering are provided.

\subsection{The nuclear charge distribution and the phase-shift analysis}

For collision energies up to a few hundred MeV, the Fourier-Bessel representation of the ground-state charge distribution,
obtained from a fit to the  measured form factors in elastic electron scattering, is commonly in use \cite{DeV}. For higher energies a Gaussian parametrization 
 of the ground-state charge distribution is required, which 
has weaker oscillations near the origin and is given by \cite{DeV}
$$\varrho_N(r) \,=\,\frac{Z}{2\pi^{3/2}d^3}\sum_{i=1}^n
\frac{Q_i}{1+2(R_i/d)^2}$$
\begin{equation}\label{3.1}
\times \;\left[ e^{-(r-R_i)^2/d^2}+\,e^{-(r+R_i)^2/d^2}\right],
\end{equation}
where $Q_i,R_i,d$ and $n$ are fit parameters.
This Gaussian parametrization is introduced into the ELSEPA code \cite{SJ05},
which 
  is an update and extension  of the Fortran package RADIAL \cite{Sal} for calculating the electronic scattering states from the radial Dirac equation.
It was recently applied at GeV energies by Koshchii et al \cite{Ko21}.

Given $\varrho_N$ from (\ref{3.1}), the nuclear potential $V_T$ is generated by means of
\begin{equation}\label{3.2}
V_T(r)\,=\,-4\pi\left[ \frac{1}{r}\int_0^r r'^2dr'\,\varrho_N(r')\,+\int_r^\infty r'dr'\,\varrho_N(r')\right],
\end{equation}
which has to be evaluated numerically.
This means that in our code for calculating the radiative corrections, which is based upon the RADIAL package, the potential $V_T$ from (\ref{3.2}) is substituted 
for the previously used analytical potential derived from  the Fourier-Bessel representation of $\varrho_N$.
Moreover, for an efficient way of performing the partial-wave analysis for potential scattering, the subroutine DFREE of the RADIAL package is modified by moving its grid-merging part into a separate subroutine
which has to be called prior to the performance of the sum over the partial waves. 
In order to allow for very small scattering angles
($\vartheta_f \approx 5^\circ$) the convergence acceleration for the partial-wave sum \cite{YRW} is reduced from
three-fold to two-fold
as done in \cite{Ko21}. Up to 25000 partial waves are considered to provide sufficient accuracy.

For minimizing numerical instabilities in the Sherman function tiny precision limits are required. These pertain to the limit of the inner phase shift (due to the short-range part of $V_T:\;|\delta_{\kappa}| < 10^{-12}$ for sufficiently large angular momenta $\kappa$),
implying that for larger angular momenta the Dirac partial waves are replaced by the point-Coulomb ones.
They also concern the accuracy of the partial-wave sums ($\epsilon < 10^{-10}-10^{-12}$, depending on collision energy and scattering angle,
where $\epsilon$ is the difference between the partial-wave sums relating to $\kappa$ and $\kappa +1$, respectively).

Fig.2a compares the Coulombic Sherman function $S_{\rm coul}$ for $^{208}$Pb at two forward scattering angles,
$5^\circ$ and $25^\circ$, when calculated from the Fourier-Bessel and the Gaussian charge distribution, respectively.
While at the foremost angle there is no visible difference between the two representations of $\varrho_N(r)$ up to 1 GeV, at $25^\circ$ the deviations start
already near 500 MeV.
In the angular dependence of $S_{\rm coul}$ (Fig.2b) the Fourier-Bessel representation fails completely at angles above $25^\circ$ when the collision energy
 reaches the GeV region, while only the Gaussian charge density guarantees a regular diffraction pattern.
For the small $^{12}$C nucleus (Fig.2c), the difference between the two prescriptions starts only at angles above $40^\circ$ (for 950 MeV), respectively above $60^\circ$ (for 570 MeV).

\begin{figure}
\vspace{-1.5cm}
\includegraphics[width=11cm]{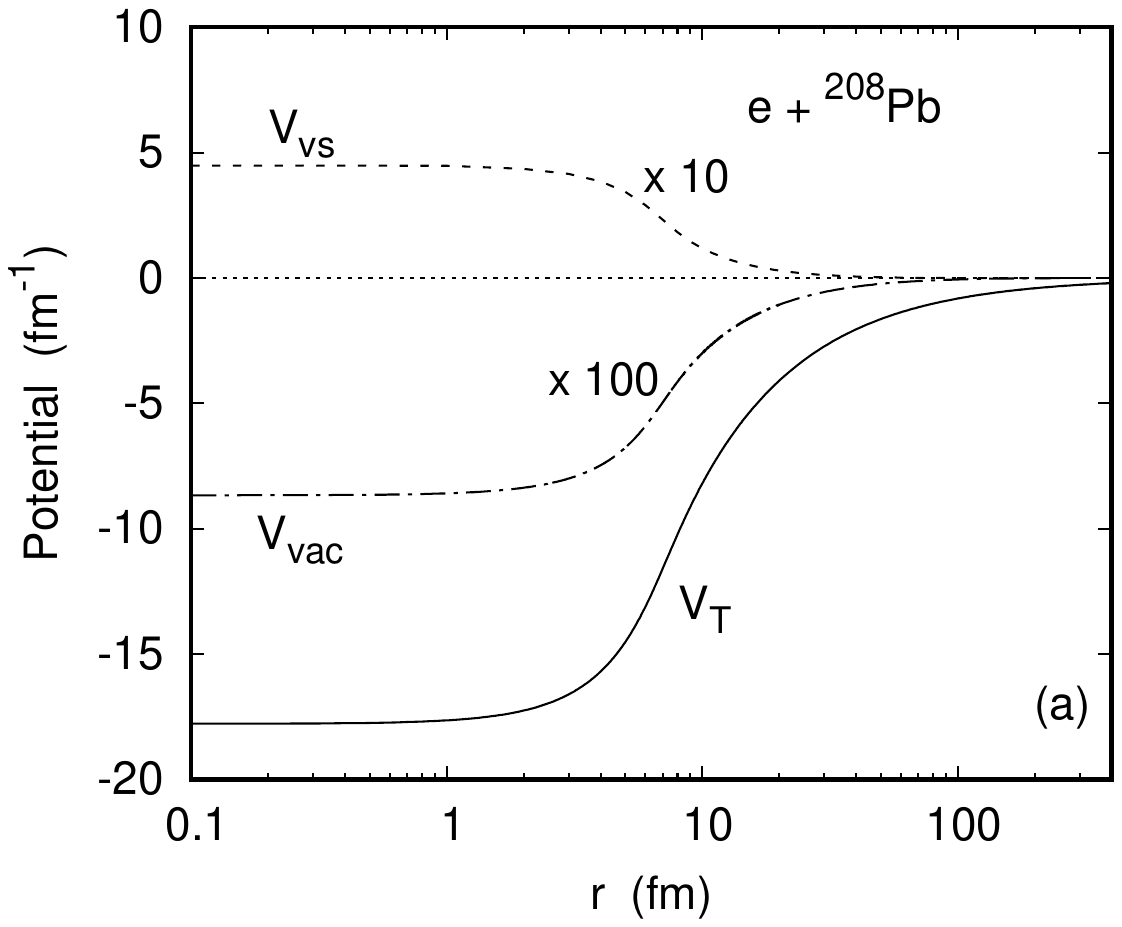}
\vspace{-1.5cm}
\vspace{-0.5cm}
\includegraphics[width=11cm]{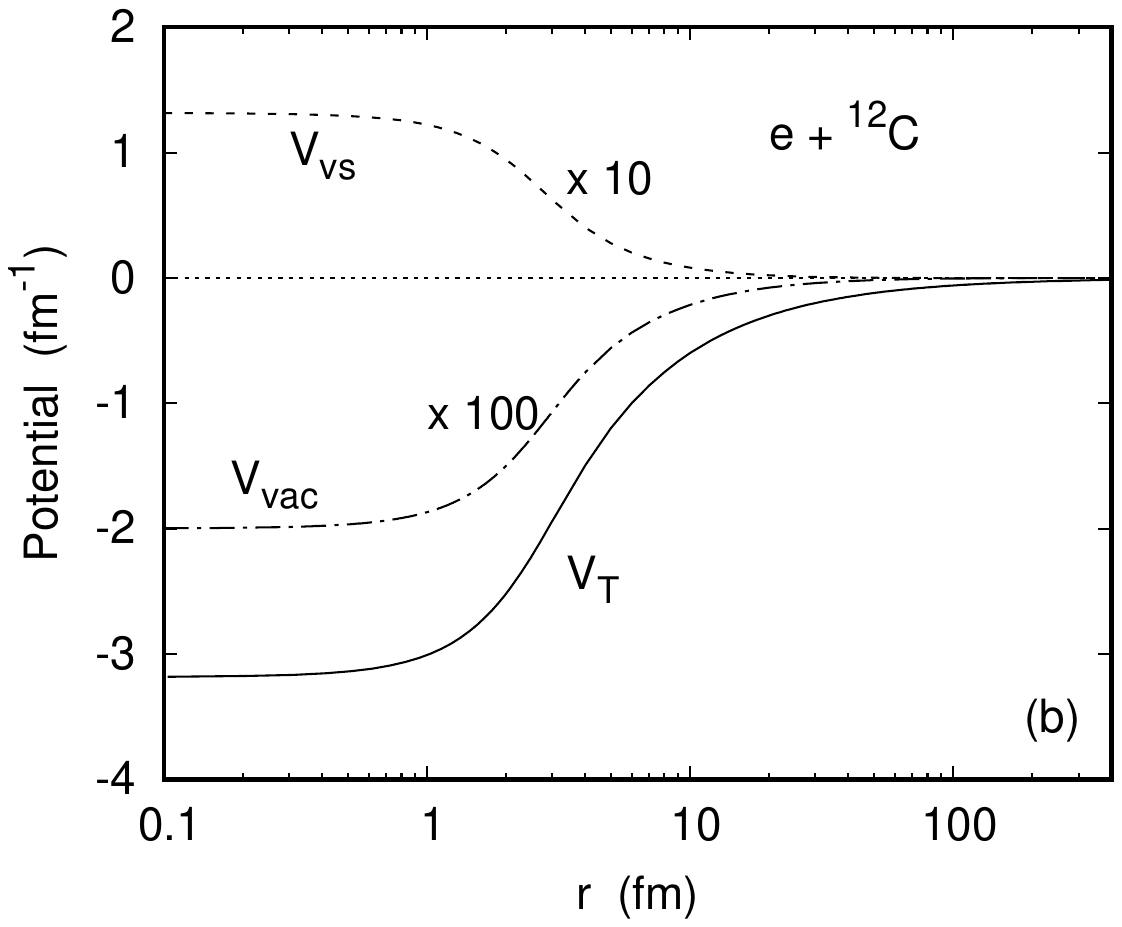}
\caption
{
Spatial dependence of the Uehling potential $V_{\rm vac}\;(-\cdot -\cdot -)$ and of the vertex plus self-energy potential $V_{\rm vs}\;(----)$ (a) for $^{208}$Pb and (b) for $^{12}$C.
Included is the Coulombic potential $V_T$ (-------).
For better visibility $V_{\rm vs}$ is multiplied by a factor of 10 and $V_{\rm vac}$ by a factor of 100.
}
\end{figure}

\subsection{The QED potentials}

From Fig.3 it follows that the potentials for vacuum polarization and for the vertex plus self-energy correction exhibit a steep fall-off beyond the nuclear radius $R_N$.
However, their radial dependence at large $r$ differs considerably from that of the Coulombic field, such that the sum of the three
potentials shows a non-Coulombic behaviour even at distances extending to more than 1000 fm.
Since the long-range part of the QED potentials becomes increasingly important when the collision energy gets larger,
one might think of extending the matching point between the inner (Dirac) and outer (point-Coulomb) radial solutions of the Dirac equation to such large values of $r$.
However, this leads not only to excessively long computation times but also to unphysical structures in the Sherman function,
since the accuracy of $V_{\rm vs}$ as calculated from (\ref{2.11}) deteriorates for large $r$.
Taken into consideration that at such distances the absolute value of $V_{\rm vs}$ has decreased by many orders of magnitude (Fig.4),
an exponential tail is fitted at $\tilde{r}=400$ fm such that the modified potential reads
\begin{equation}\label{3.3}
V_{\rm vs}^{\rm mod}(r)\,=\,\left\{ \begin{array}{cc}
V_{\rm vs}(r),& r <\tilde{r}\\
V_{\rm vs}(r)[1+\lambda(r-\tilde{r})]e^{-\lambda(r-\tilde{r})},& \tilde{r} \leq r < r_{\rm cut}\\
0,& r \geq r_{\rm cut}
\end{array} \right.
\end{equation}
with $\lambda=10^{-2}$/fm and $r_{\rm cut}\approx 900$ fm.
The tail is chosen in such a way that in $r=\tilde{r},\;V_{\rm vs}^{\rm mod}$ is continuous and differentiable.
It should be kept in mind that even with this exponential tail the number of integration steps in the $q$-integral (\ref{2.11}) has to be taken very large ($\sim 4000$)  in order to achieve a monotonous $r$-dependence of $V^{\rm mod}_{\rm vs}$ beyond 400 fm.
The vs potential above 100 fm with and without exponential tail is displayed in Fig.4.

\begin{figure}
\vspace{-1.5cm}
\includegraphics[width=11cm]{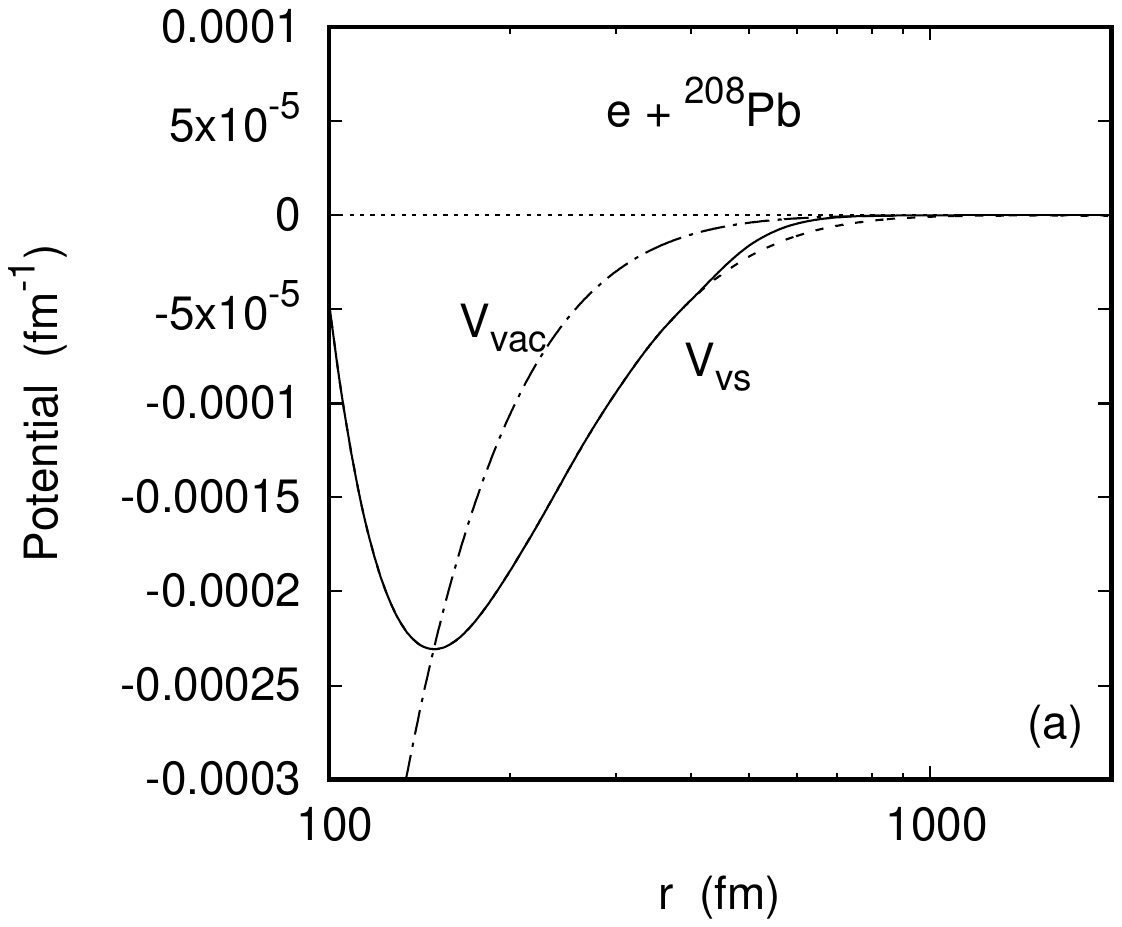}
\vspace{-1.5cm}
\vspace{-0.5cm}
\includegraphics[width=11cm]{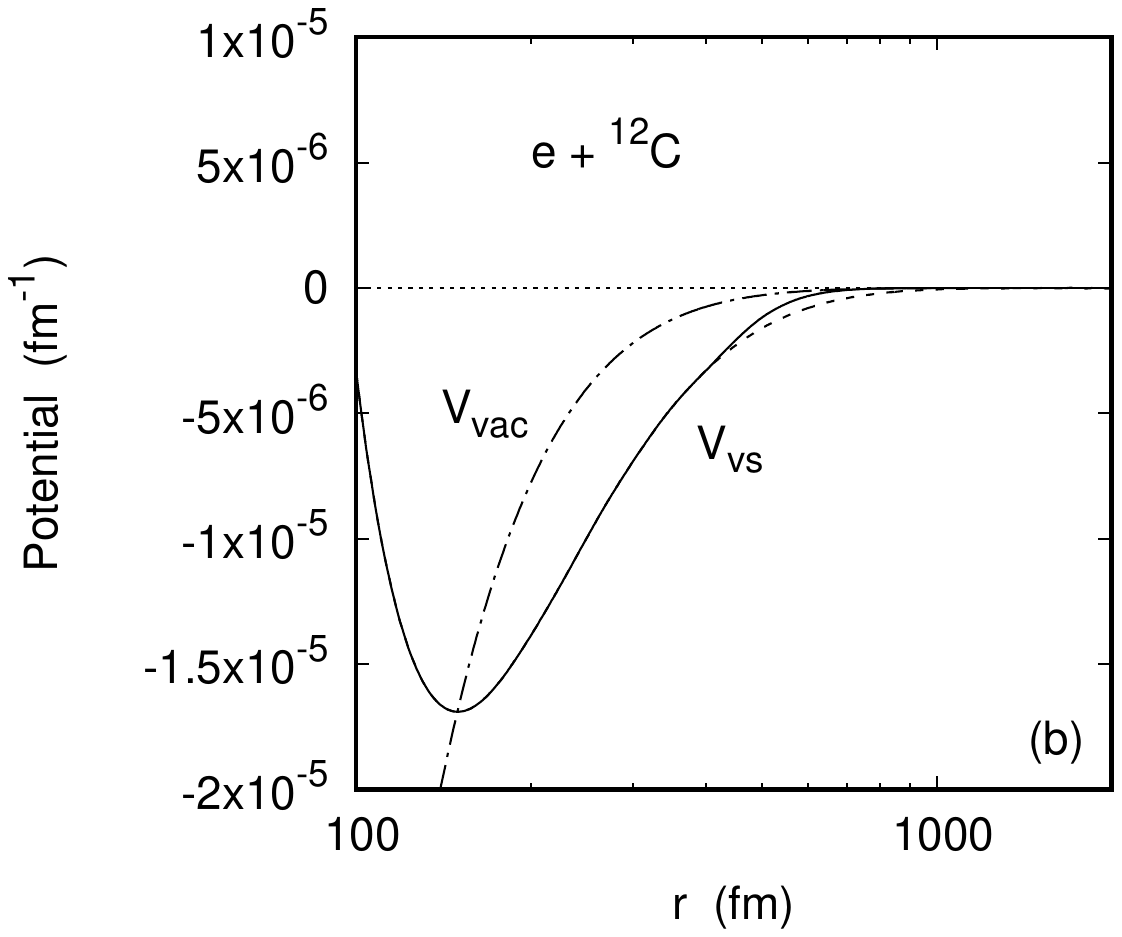}
\caption
{
Large-distance behaviour of $V_{\rm vac}\;(-\cdot -\cdot -)$ and of $V_{\rm vs}\;(----)$ (a) for $^{208}$Pb and (b) for $^{12}$C.
Also shown is $V_{\rm vs}^{\rm mod}$ with the fitted tail (--------).
}
\end{figure}

For the Uehling potential, which is monotonous in $r$ even at very large distances, the asymptotic formula (\ref{2.9}) is used for $r>800$ fm (which corresponds to about 100 $R_N$) and integrated up to 1500 fm.
From Figs.3 and 4 it is seen that $V_{\rm vac}$ is negative for all $r$, whereas $V_{\rm vs}$ changes sign near 96 fm.

Our improved code was tested against the previous code for collision energies up to 150 MeV (the energy region covered in \cite{Jaku24}).
No difference was found for the cross section change, while there occurred no longer wiggles in the carbon spin asymmetry change at the higher energies or at the larger angles. However, even the improved code does not allow for accurate predictions of the QED-modified Sherman function at energies in the GeV region.

\begin{figure}
\vspace{-1.5cm}
\includegraphics[width=11cm]{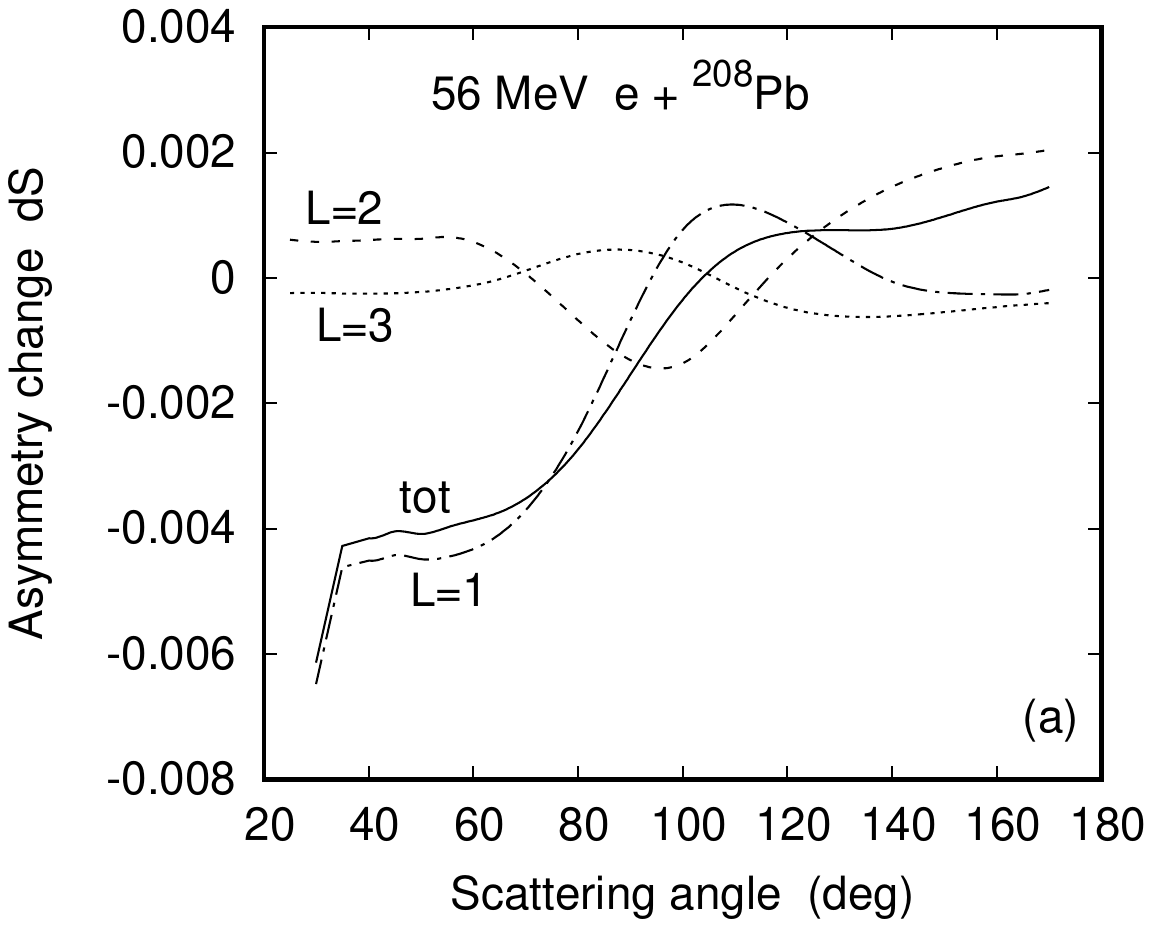}
\vspace{-1.5cm}
\vspace{-0.5cm}
\includegraphics[width=11cm]{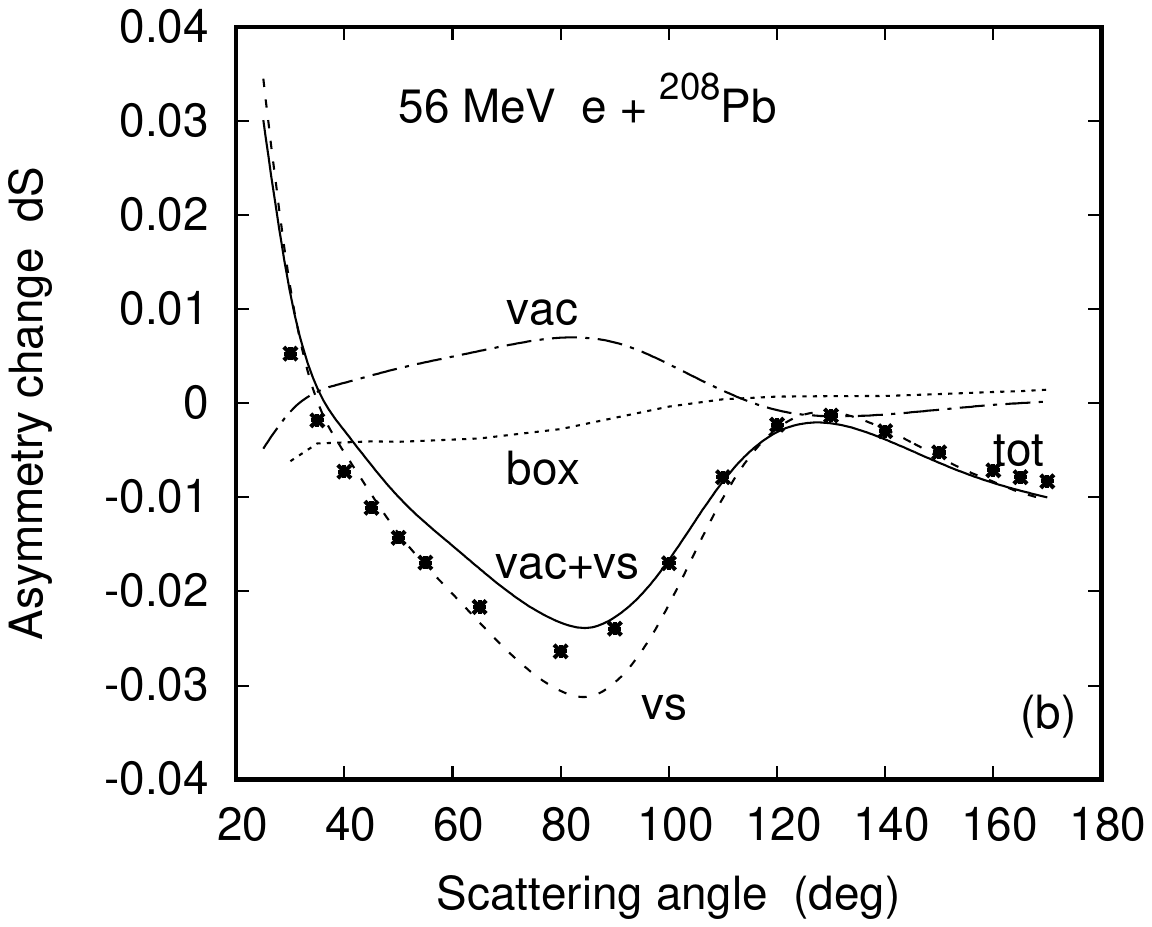}
\caption
{
Relative change $dS$ of the spin asymmetry from 56 MeV $e + ^{208}$Pb collisions as a function of scattering angle $\vartheta_f$.
In (a), the dispersive change $dS_{\rm box}$ is shown (-------), together with its contributions from all dipole states $(-\cdot -\cdot-)$, all quadrupole states $(----)$ and all octupole states $(\cdots\cdots)$.
(b) shows the QED changes (---------), resulting from vacuum polarization $(-\cdot -\cdot -)$ and the vs correction $(----)$.
Also shown is the dispersive change $dS_{\rm box}$ from (a) $(\cdots\cdots)$. The sum of all radiative changes is included $(\ast \ast \ast)$.
(Updating Fig.8b in \cite{Jaku24}.)
}
\end{figure}

\section{Results}
\setcounter{equation}{0}

In this section we present the results for the dispersive modification of the spin asymmetry by nuclear excitations up to 30 MeV,
 as well as estimates for the QED corrections to the Sherman function.
Two targets, the $^{208}$Pb nucleus and the $^{12}$C nucleus, are considered.

\subsection{The $^{208}$Pb nucleus}

For $^{208}$Pb, the  five dipole states at $\omega_L=14.2$, 12.3, 14.6, 15 and 5.512 MeV (listed according to their importance) were taken into account
for dispersion. In addition, three quadrupole states
(at 10.9, 21.6 and 4.085 MeV) as well as two octupole states (at 2.515 and 28.94 MeV) were considered.

The angular distribution of the relative spin asymmetry change for 56 MeV electrons is displayed in Fig.5.
While in the forward hemisphere the dipole states provide the dominant contribution,
the quadrupole excitations gain importance at large scattering angles.
Due to a partial cancellation between the $L=1$ and $L=2$ contributions, even the octupole states cannot be neglected beyond $100^\circ$ (Fig.5a).
The modification of the spin asymmetry by the QED effects (Fig.5b) exceeds the dispersion effects.
Nevertheless, both effects have to be taken into account at the smaller angles according to (\ref{2.26}) ($\Delta\sigma_{\rm QED}$ varies with $\vartheta_f$ from 4\% to 15\%).

\begin{figure}
\vspace{-1.5cm}
\includegraphics[width=11cm]{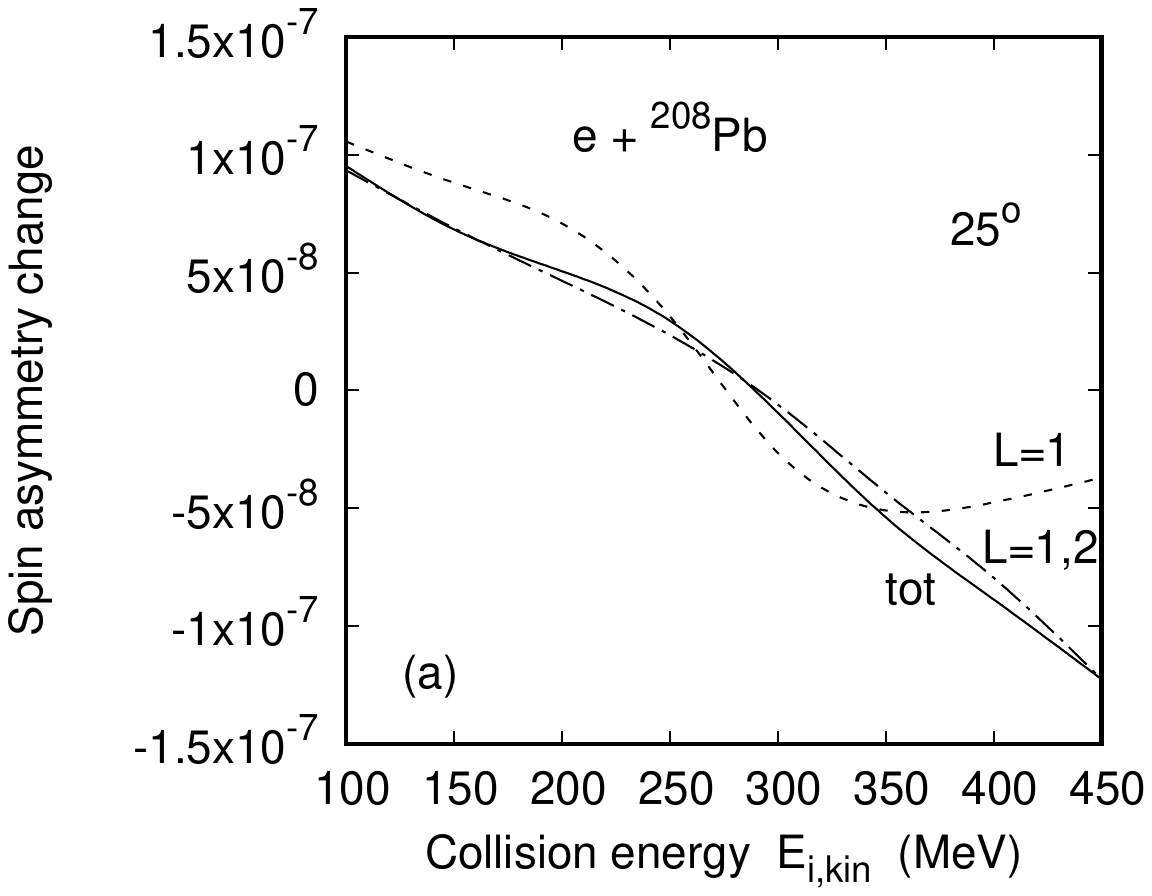}
\vspace{-1.5cm}
\vspace{-0.5cm}
\includegraphics[width=11cm]{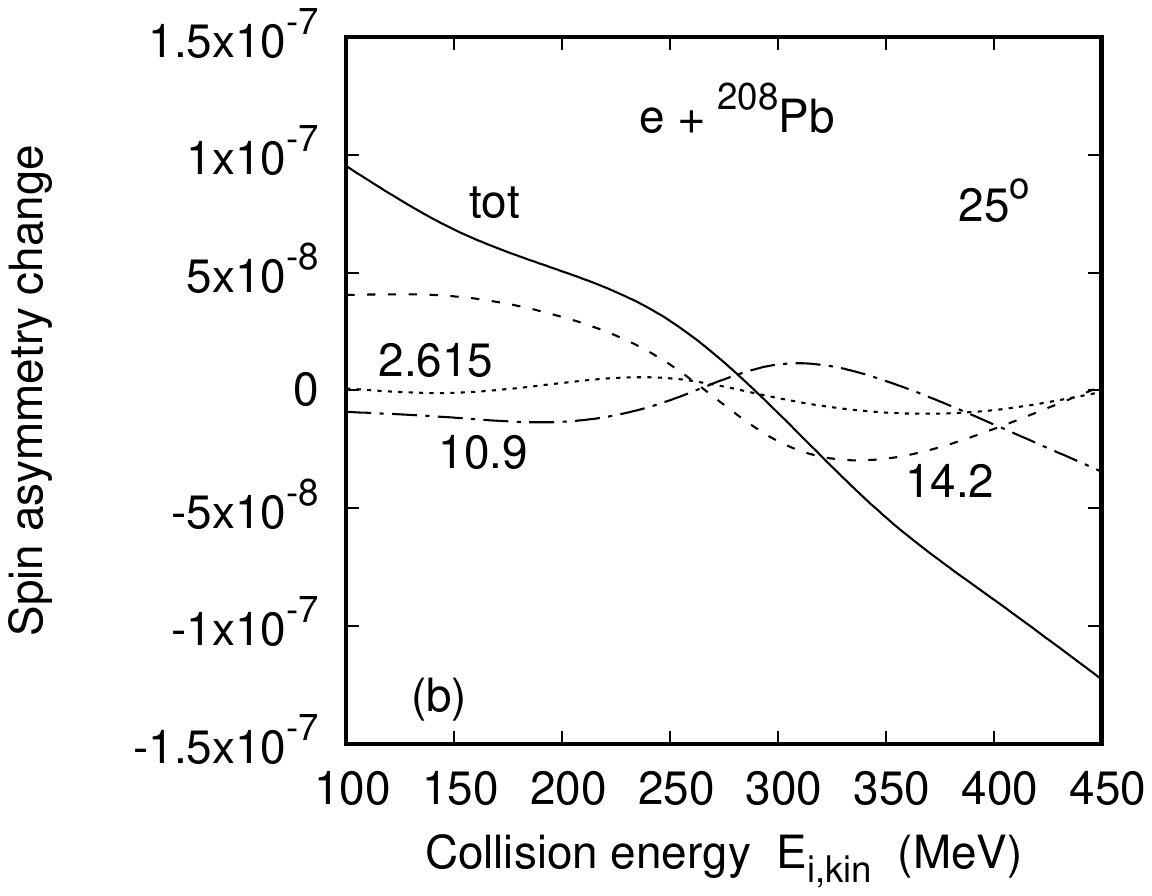}
\vspace{-1.5cm}
\vspace{-0.5cm}
\includegraphics[width=11cm]{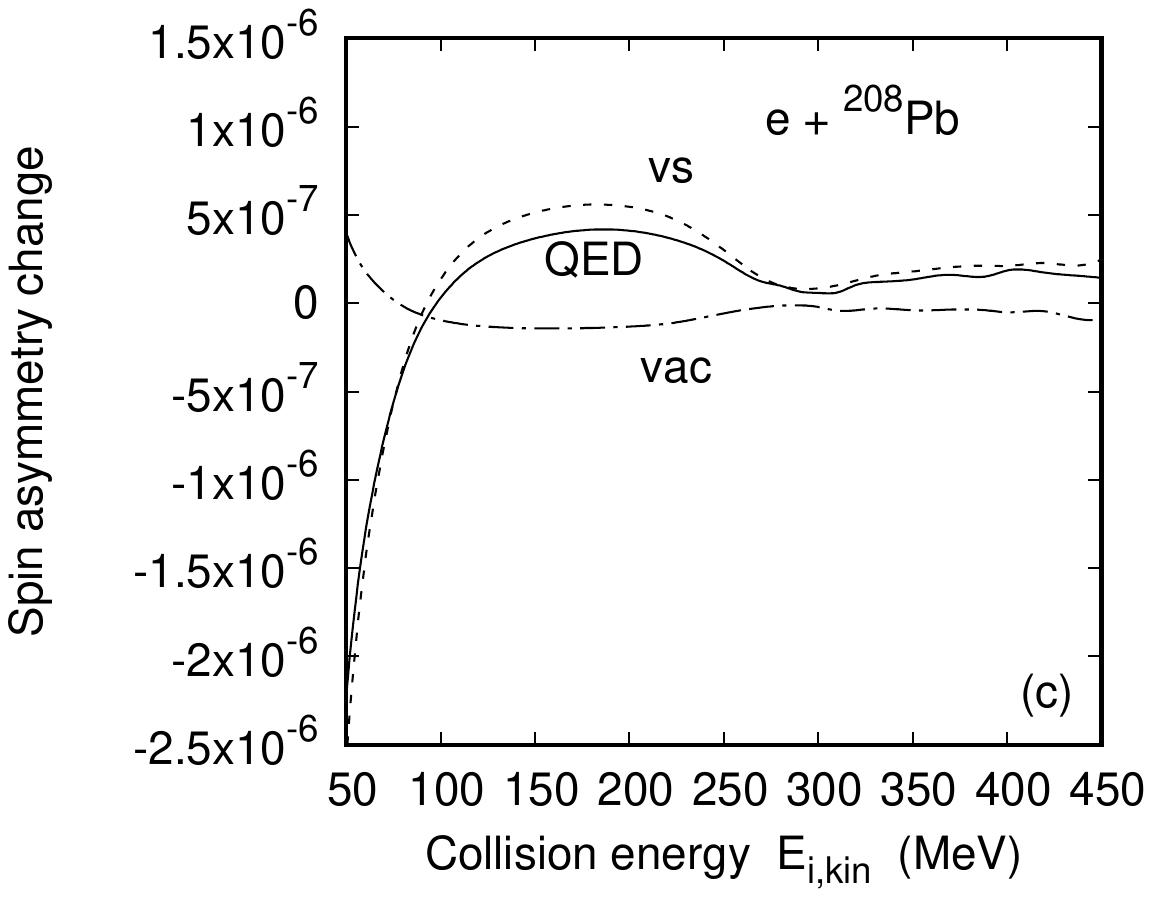}
\caption
{
Energy distribution of the spin asymmetry change in $e+^{208}$Pb collisions at $\vartheta_f=25^\circ$.
(a) Total change $S_{\rm box} -S_{\rm coul}$ by dispersion (-------), its dipole contribution $(----)$
and the sum of dipole and quadrupole contributions $(-\cdot - \cdot -)$. 
(b) Separate dispersive contributions from the 14.2 MeV dipole state $(-----)$, the 10.9 MeV quadrupole state $(-\cdot - \cdot -)$, the 2.615 MeV octupole state $(\cdots\cdots)$ and the total change $S_{\rm box} -S_{\rm coul} $ (--------).
(c) QED contributions from vacuum polarization $(-\cdot-\cdot -)$ and from the vs correction $(----)$ together with their sum $S_{\rm QED} -S_{\rm coul}$ (----------).
}
\end{figure}

The energy dependence of the absolute spin asymmetry change at an angle of $25^\circ$  is shown in Fig.6.
It is seen that at all energies above 100 MeV, the quadrupole excitations cannot be neglected, but they become particularly important at the highest energies (Fig.6a).
The $L=3$ states are of minor importance as can also be inferred from Fig.6b where the results from the dominant $L=1,2,3$ excitations are shown separately.

Fig.6c depicts the energy dependence of the  QED corrections,  the  vacuum polarization and the vs effect.
At high collision energies the vacuum polarization and the vs correction are in general of opposite sign, and their ratio is often considerably larger than the factor of 2.5 obtained from bound-state considerations \cite{Sh00}, supporting
the findings at the lower energies \cite{Jaku24}.
The accuracy of the QED corrections, particularly of vacuum polarization, deteriorates at this angle for energies above 250 MeV, and the wiggles are caused by convergence problems.
Note the large dominance of the QED effects below 250 MeV as compared to dispersion.

\subsection{The $^{12}$C nucleus}

For the $^{12}$C target the following excited states were considered: Two $1^-$ states at $\omega_L=23.5$ and 17.7 MeV, two $2^+$ states at 4.439 and 9.84 MeV
and the $3^-$ states at 9.64 and 14.8 MeV.
It turned out, however, that the octupole states give no contribution for any collision energy or angle considered.

We start by investigating the behaviour of the dispersive spin asymmetry at small scattering angles.
The hadronic model predicts a linear increase with $\vartheta_f$ according to $\sin (\vartheta_f/2)$ \cite{AM04,BK06,Ko21},
in contrast to the cubic increase of $S_{\rm coul}$ from potential scattering or from the Friar-Rosen theory for dispersion \cite{FR74,Jaku23}.

In order to study the angular dependence at angles close to zero where the phase-shift analysis fails (because the necessary convergence acceleration generates a singularity at $\vartheta_f=0)$,
the first-order Born approximation can be used for light targets.
To lowest order in $A_{fi}^{\rm box}$, the dispersive spin asymmetry, neglecting Coulomb distortion, is calculated from

$$S_{\rm box}^{\rm Born} \,= \;\lim_{Z \to 0} (S_{\rm box} -S_{\rm coul})$$
\begin{equation}\label{4.1}
=\;\frac{\sum_{\sigma_f}2\mbox{ Re }\{ A_{fi}^{\ast B1}A_{fi}^{\rm box} \} (\uparrow -\downarrow)}{2\sum_{\sigma_f} |A_{fi}^{B1}|^2}.
\end{equation}

\begin{figure}
\vspace{-1.5cm}
\includegraphics[width=11cm]{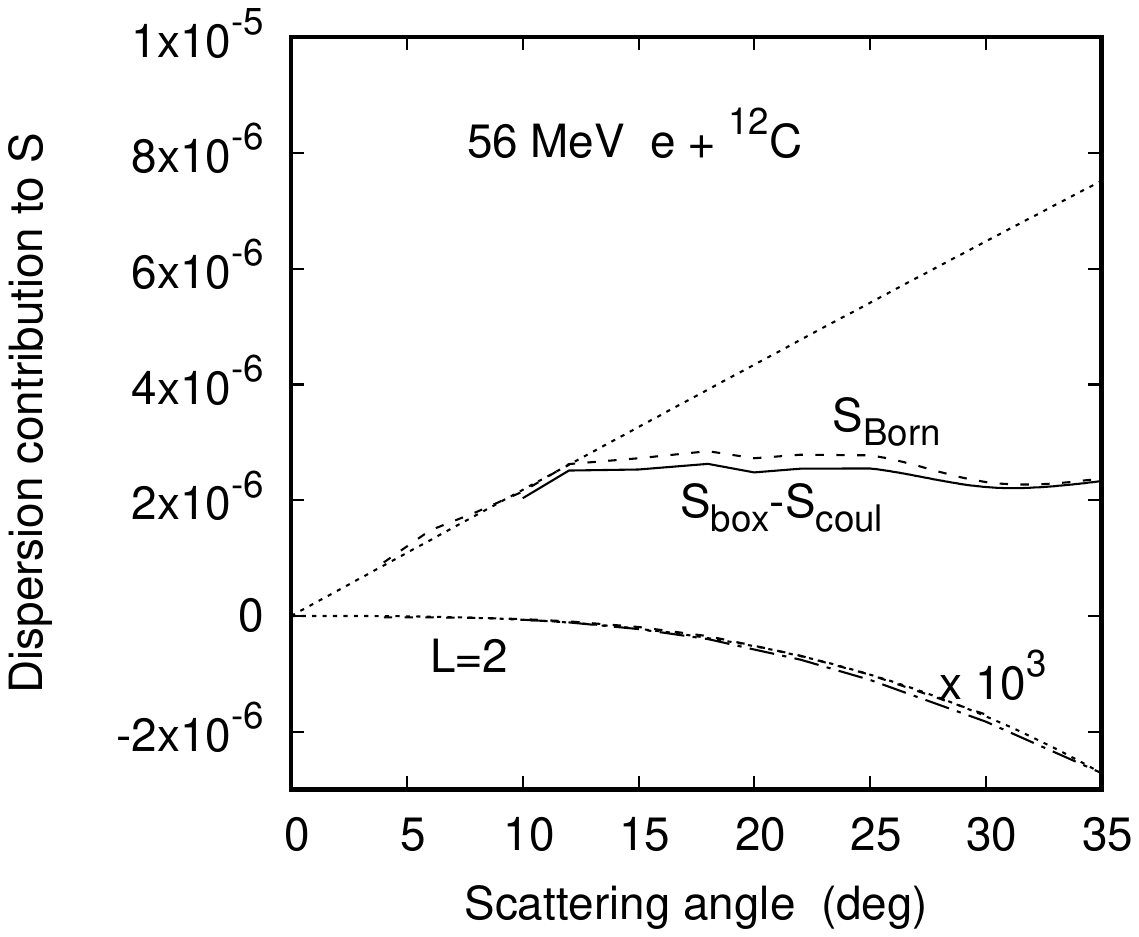}
\caption
{Influence of dispersion on the spin asymmetry from 56 MeV $e + ^{12}$C collisions as a  function of scattering angle $\vartheta_f$. Shown is $S_{\rm box}-S_{\rm coul}$ from (\ref{2.7}) (----) and $S_{\rm box}^{\rm Born}$ from (\ref{4.1}) ($----$, upper curve).
Included are the separate contributions from the $L=2$ states ($-\cdot -\cdot -$) together with their Born approximation ($----$, lower curve), both multiplied by a factor of $10^3$.
Also shown are the $\sin (\vartheta_f/2)\;(\cdots\cdots$, upper curve) and the $\sin^3(\vartheta_f/2) \;(\cdots\cdots$, lower curve) dependencies. 
The wiggles in the dipole results are due to numerics. (Updating Fig.8 in \cite{Jaku23}.)
}
\end{figure}

The angular distribution in Fig.7 for 56 MeV collision energy demonstrates the $\sin(\vartheta_f/2)$-behaviour for $\vartheta_f \to 0$ within the Born theory.
Our results for $\vartheta_f \geq 10^\circ$, which account for Coulomb distortion, are close to the Born results (with deviations between $5-10\%$ for $\vartheta_f \lesssim 30^\circ$). Thus it is conjectured that the linear increase holds also for the more accurate theory,
\begin{equation}\label{4.2}
S_{\rm box} -S_{\rm coul} \,\approx\,S_{\rm box}^{\rm Born} \,\sim\,\sin (\vartheta_f/2),\qquad \vartheta_f \to 0.
\end{equation}
The angular region of linearity shrinks with energy as is demonstrated in \cite{Ko21} for GeV impact energies.

In contrast, the $2^+$ excitations at 4.439 MeV  and at 9.84 MeV (with nearly equal results, their sum being
included in the figure) exhibit a cubic increase ($\sim \sin^3(\vartheta_f/2)$), but these states do not contribute at forward angles.

\begin{figure}
\vspace{-1.5cm}
\includegraphics[width=11cm]{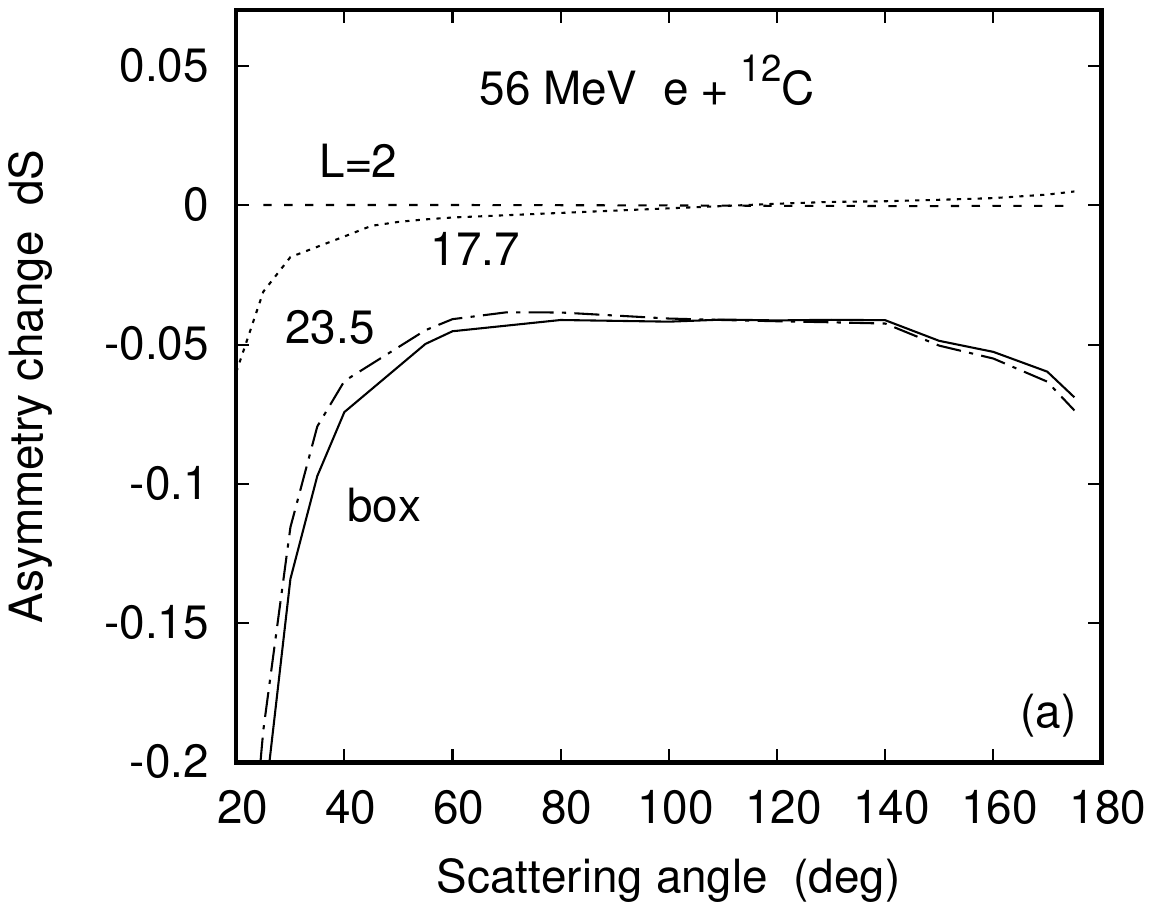}
\vspace{-1.5cm}
\vspace{-0.5cm}
\includegraphics[width=11cm]{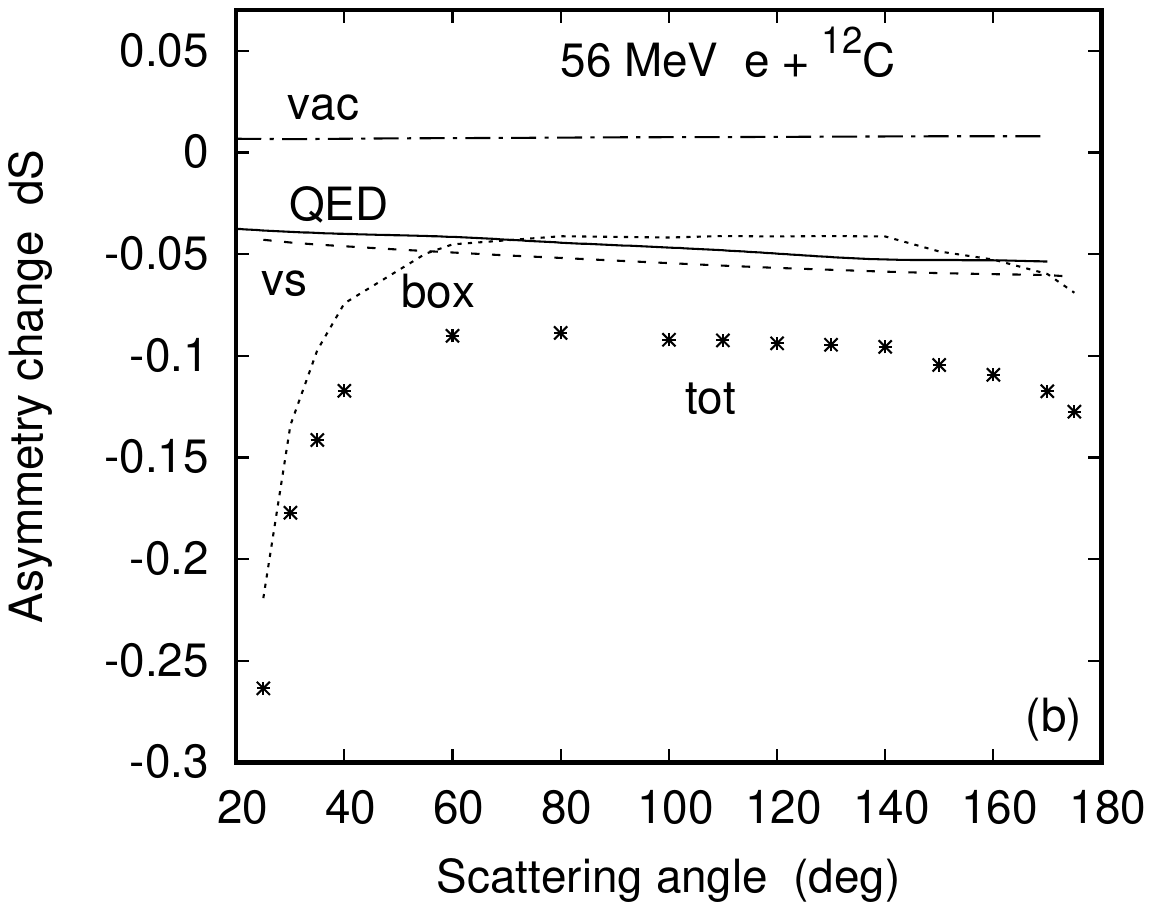}
\caption
{
Relative change $dS$ of the spin asymmetry from 56 MeV $e + ^{12}$C collisions as a function of scattering angle $\vartheta_f$. In (a) the dispersive change $dS_{\rm box}$ is shown (------), together with its constituents from the 23.5 MeV $(-\cdot -\cdot -$) and 17.7 MeV ($\cdots\cdots$) dipole states and from the sum of the two quadrupole states $(----$).
The structures from the logarithmic singularity in the differential cross section (at $q\approx 2 \omega_L$) are avoided by selecting appropriate grid points. (Updating Fig.6a in \cite{Jaku23}.)\\
(b) shows the QED changes (-------), resulting from vacuum polarization $(-\cdot -\cdot -$) and the vs correction $(----$). Also shown is  the dispersive change $dS_{\rm box}$ from (a) $(\cdots\cdots)$.
The sum of all radiative changes is included $(\ast\ast\ast)$. (Updating Fig.8a in \cite{Jaku24}.)
}
\end{figure}

The angular distribution of the relative spin asymmetry change at larger angles is displayed in Fig.8.
Clearly, the dominant contribution results from the highest dipole state considered, at 23.5 MeV, while the quadrupole contribution is unimportant at all  angles (Fig.8a).

In Fig.8b the relative changes from vacuum polarization
and the vs correction are shown in comparison with those from dispersion.
While vacuum polarization gives a small contribution of about 1\%, the QED corrections,  basically due to the vs process, amout to 5\%, being  nearly independent of scattering angle.
Beyond $60^\circ$ the QED and dispersive corrections are of similar magnitude, while at the smaller angles $dS_{\rm box}$ is largely dominating.
The strong increase of $|dS_{\rm box}|$ with decreasing angle  for $\vartheta_f < 40^\circ$ is basically due to the fact that $S_{\rm coul} \to 0$ for $\vartheta_f \to 0$ (cf. (\ref{2.7a})).

Also shown is the superposition of both effects according to (\ref{2.26}), which leads to a correction of the beam-normal spin asymmetry by about 10\% at the higher angles.
Note that a simple addition of the results from the two radiative processes is also here prohibited by the formidable size of the QED cross section change ($\Delta\sigma_{\rm QED}$ decreases from -0.034 at $25^\circ$ to -0.085 at $175^\circ$, while the cross-section change by dispersion is well below $10^{-3}$).

\begin{figure}
\vspace{-1.5cm}
\includegraphics[width=11cm]{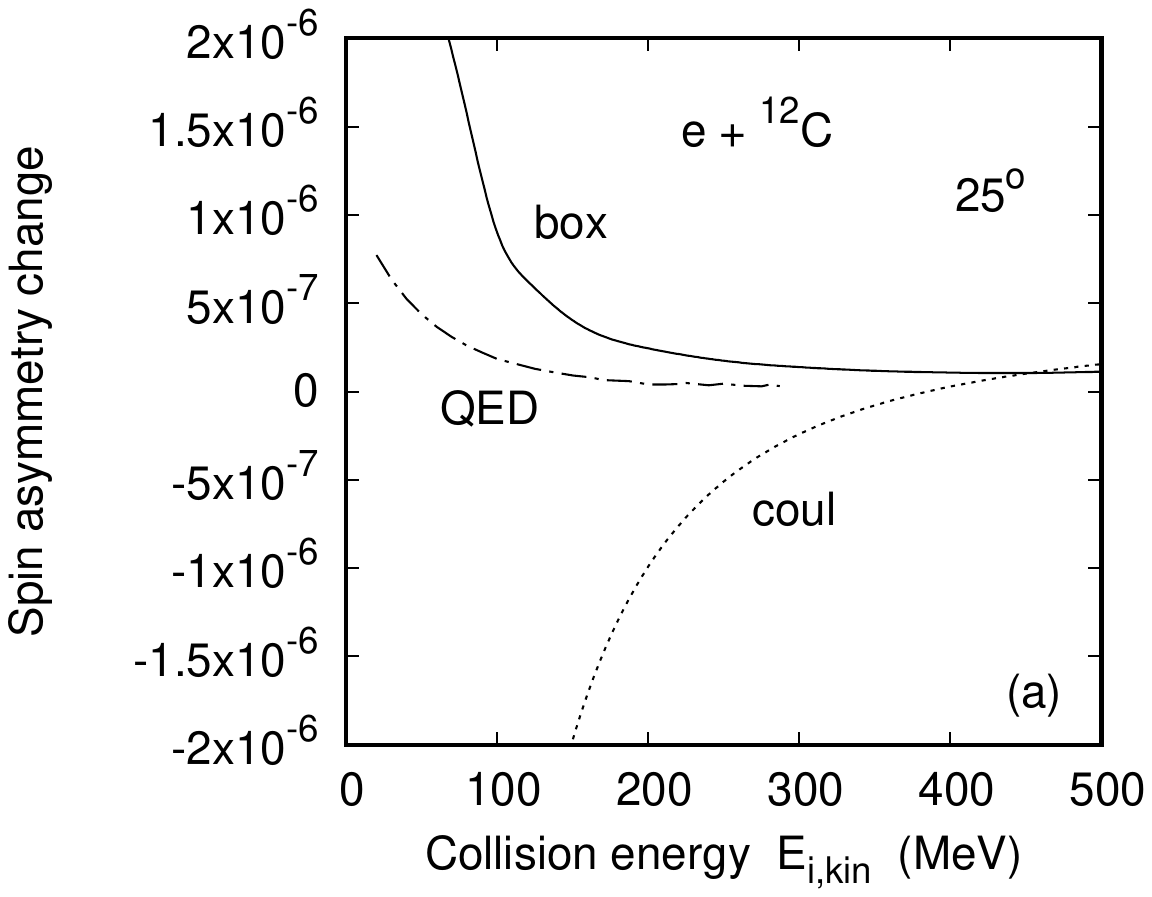}
\vspace{-1.5cm}
\vspace{-0.5cm}
\includegraphics[width=11cm]{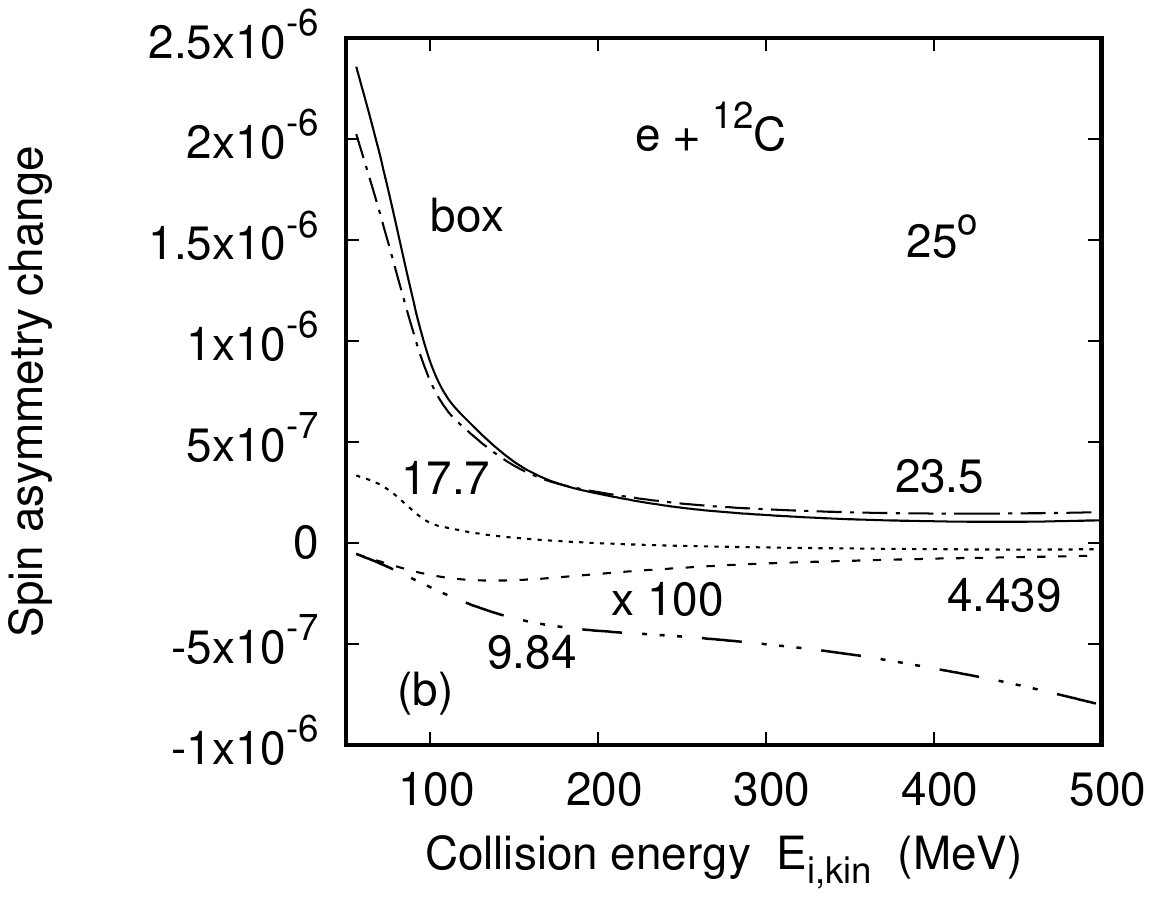}
\vspace{-1.5cm}
\vspace{-0.5cm}
\includegraphics[width=11cm]{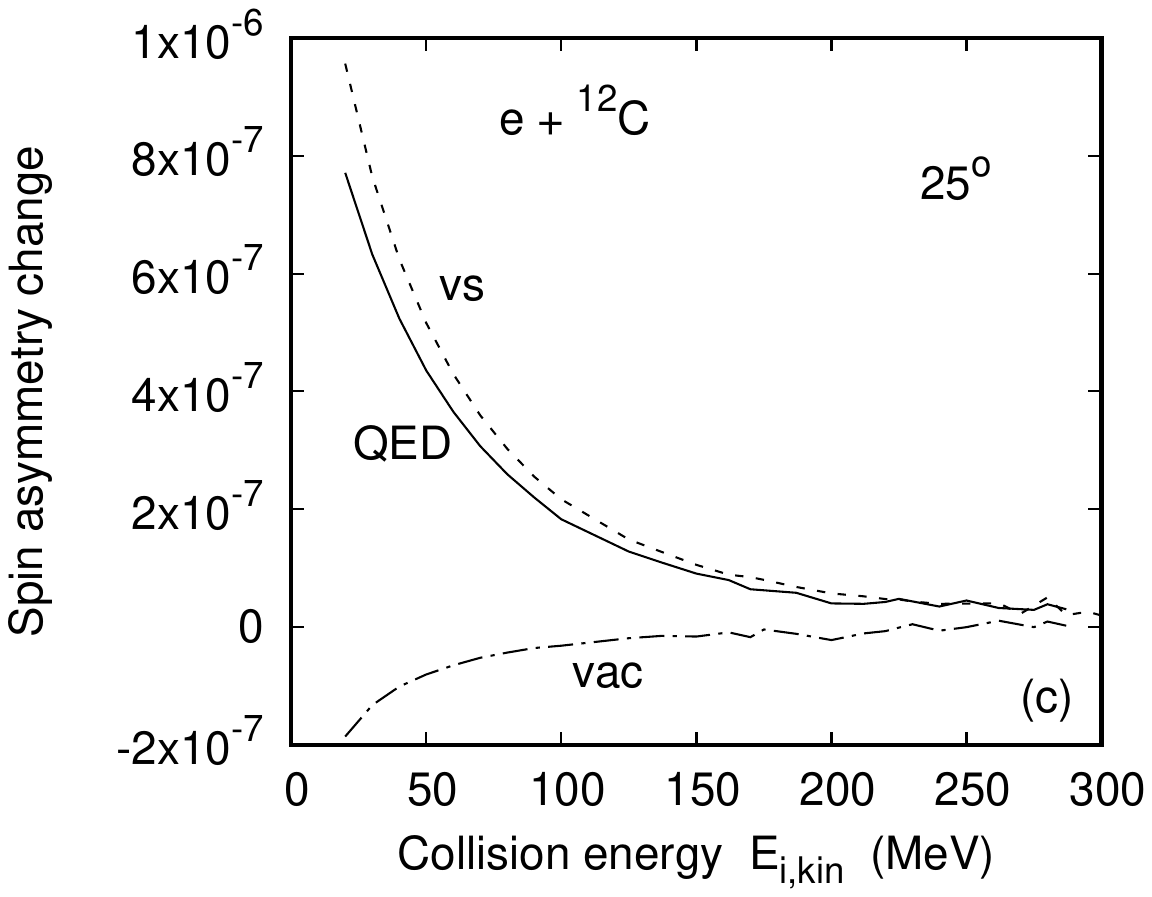}
\caption
{
Energy distribution of the spin asymmetry change in $e+^{12}$C collisions at $\vartheta_f=25^\circ$.
(a) Total change $S_{\rm box} -S_{\rm coul}$ by dispersion (-------), the change $S_{\rm QED} -S_{\rm coul}$ by vacuum polarization and the vs correction $(-\cdot -\cdot -)$ as well as $S_{\rm coul} \;(\cdots\cdots)$.
(b) Separate dispersive contributions from the 23.5 MeV  $(-\cdot -\cdot -)$ and the 17.7 MeV  $(\cdots\cdots)$ dipole states, as well as from the 4.439 MeV ($----$) and 9.84 MeV $(-\cdots-$)  quadrupole states,
both multiplied by a factor of $10^2$. The total change from (a) is
included (--------).
(c) Separate QED contributions from vacuum polarization $(-\cdot-\cdot -)$ and from the vs correction $(----)$ together with their sum $S_{\rm QED} -S_{\rm coul}$ (----------).
}
\end{figure}

Next we consider the energy distribution of the total spin asymmetry change. $S-S_{\rm coul}$ is shown in Fig.9a at the forward angle of $25^\circ$. 
Compared are the results from dispersion and from the QED effects with the Coulombic spin asymmetry.
As is true for the angular distribution at a fixed small energy (Fig.8),
 $S_{\rm box}-S_{\rm coul}$ is basically due to the dipole excited states.
The QED effects, being considerably smaller than dispersion, could only be estimated up to 300 MeV,
since the convergence of the spin asymmetry as calculated from (\ref{2.18}) is much poorer for $^{12}$C than for $^{208}$Pb, mostly because of its smaller absolute value.
However, our results suggest that the QED corrections remain negligibly small at the higher collision  energies.

Fig.9b displays the separate contributions to the dispersive spin asymmetry from the various excited states.
The dominant constituent arises from the 23.5 MeV dipole state,
while the quadrupole states at 4.439  and 9.84 MeV do not have much influence.

The modification of the Sherman function by  vacuum polarization and the vs correction, composing the QED effects,  is shown in Fig.9c.
Both contributions tend to zero at large energies.

\vspace{0.2cm} 

In order to investigate the situation in the backward hemisphere, a scattering angle of $170^\circ$ is chosen in Fig.10. At such angles the Coulombic spin asymmetry $S_{\rm coul}$ is very large, exceeding by far the spin asymmetry changes from the radiative processes.
The first diffraction structure of $S_{\rm coul}$   occurs already at an energy between $150-200$ MeV (Fig.10a),
which is transferred to the QED corrections but much reduced in size. 

The spin asymmetry changes from dispersion and its constituents are depicted in Fig.10b. 
These are governed by the contributions from the two dipole states for $E_{\rm i,kin} \lesssim 130 $ MeV and slightly modified by the effect of the quadrupole states at  higher energies (since their contributions largely cancel each other at this angle).
Note that the steep rise near $160^\circ$ marks the onset of the diffraction structure in $S_{\rm box} -S_{\rm coul}$.

\begin{figure}
\vspace{-1.5cm}
\includegraphics[width=11cm]{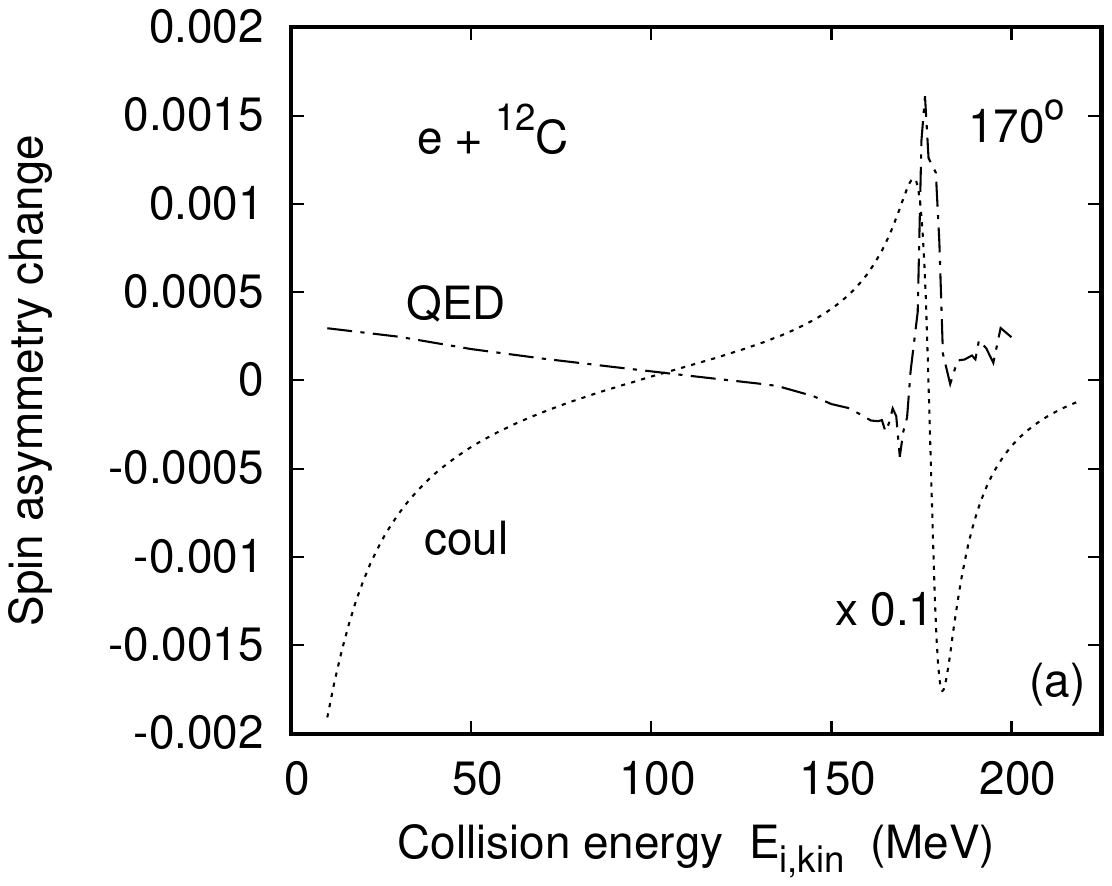}
\vspace{-1.5cm}
\vspace{-0.5cm}
\includegraphics[width=11cm]{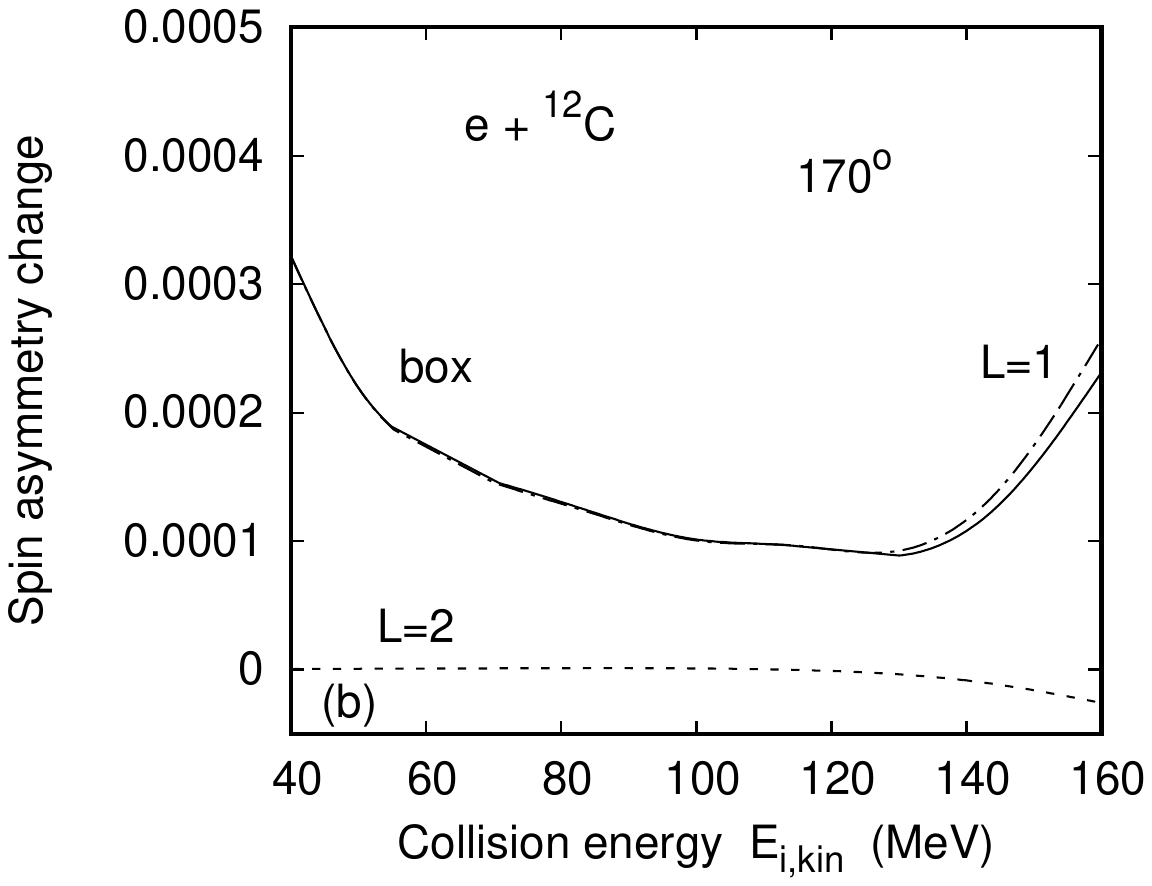}
\vspace{-1.5cm}
\vspace{-0.5cm}
\includegraphics[width=11cm]{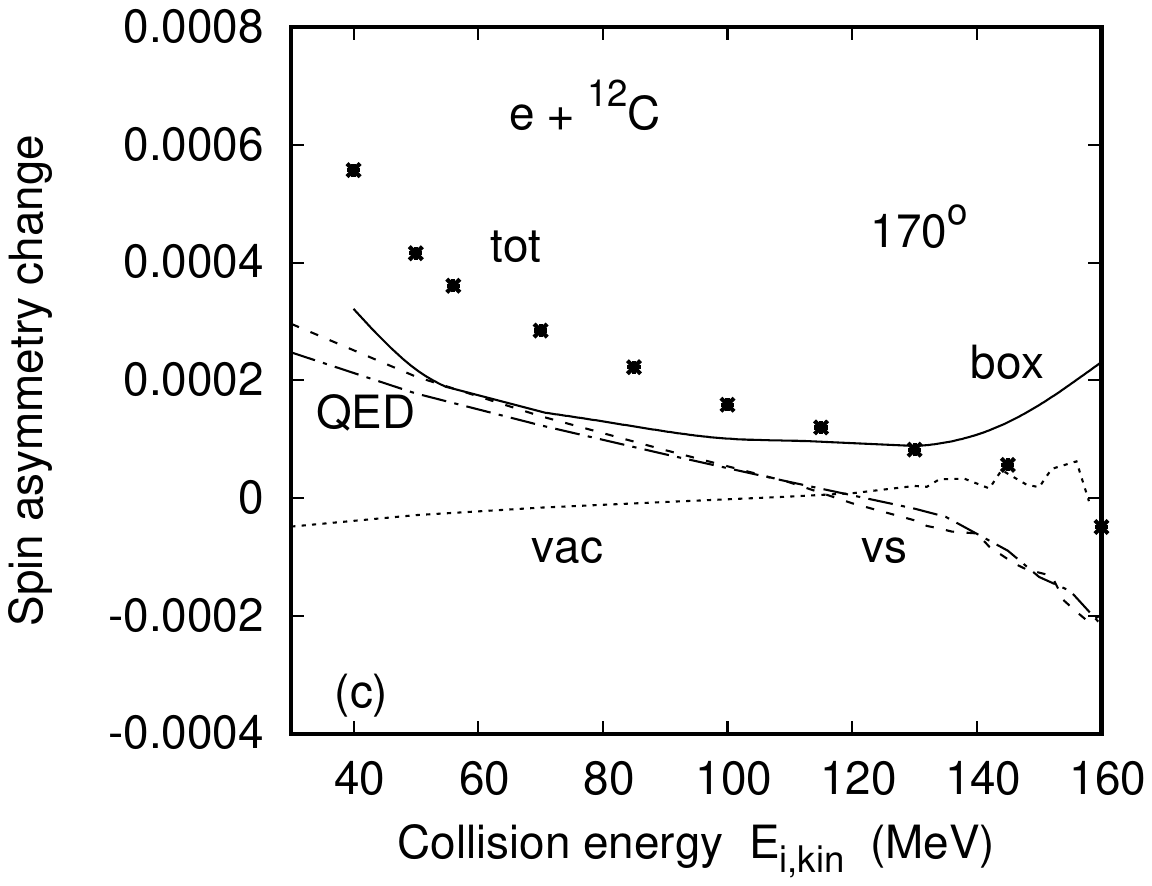}
\caption
{
Energy distribution of the spin asymmetry change in $e + ^{12}$C collisions at $\vartheta_f=170^\circ$.
(a) $S_{\rm QED} -S_{\rm coul} \;(-\cdot - \cdot -)$ in comparison with $S_{\rm coul} \;(\cdots\cdots)$ which is reduced by a factor of 0.1.
(b) Dispersion contributions from the dipole states ($-\cdot - \cdot -$), from the quadrupole states ($----$) and their sum (--------). 
(c) QED contributions from vacuum polarization $(\cdots\cdots)$ and from the vs correction $(----)$
together with their sum $(-\cdot -\cdot -)$.
Also shown is the total change $S_{\rm box}-S_{\rm coul}$ by dispersion (--------)
and the combined radiative correction $S_{\rm tot} -S_{\rm coul}$ ($\ast\ast\ast$).
The wiggles are due to numerics.
}
\end{figure}

The separate contributions from vacuum polarization and from the vs correction are shown in Fig.10c.
Like for forward angles (Fig.9c), the energy distribution of the QED corrections to the spin asymmetry is basically due to the vs process.
However, at $170^\circ$ the modifications from the QED effects increase with energy beyond 120 MeV and become  nearly comparable in size with the dispersive spin asymmetry changes.
Hence all these radiative corrections have to be considered simultaneously according to (\ref{2.24}), because of the large QED cross section changes.

\subsection{Comparison with experiment}

Finally we provide a comparison with  existing high-energy measurements for both nuclei.

For $^{12}$C there are experimental data at 570 MeV and angles between $15^\circ-26^\circ$ \cite{Es18,An21} which produce  a spin asymmetry around $S_{\rm exp} \approx -2 \times 10^{-5}$, exceeding by far the asymmetry from potential scattering ($S_{\rm coul} = 2.13 \times 10^{-7}$ at $25^\circ$).
The calculated dispersive spin asymmetry, $S_{\rm box}=1.51 \times 10^{-7}$, is too small and even differs in sign from experiment, while QED changes are also supposed to be negligible at such high energies.

There are further measurements at 1063 MeV \cite{Ab12} and at 1158 MeV \cite{An21}. For 1063 MeV and $5^\circ$, experiment reports $S_{\rm exp} = -6.5 \times 10^{-6}$. The largely dominating $L=1$ state at 23.5 MeV leads, however, only to $S_{\rm box} = 3.6 \times 10^{-8}$,
 again nearly two orders of magnitude below the measured spin  asymmetry.

For the lead nucleus there are high-energy precision measurements at 953 and 2.18 MeV \cite{Ad22} as well as earlier ones at 1063 MeV \cite{Ab12}, all around a scattering angle of $5^\circ$.
The experiment at 953 MeV and $4.7^\circ$ (where $S_{\rm coul}=-1.77 \times 10^{-7}$) reports an asymmetry of $4.0 \times 10^{-7}$. Our theoretical results are $S_{\rm box}=-1.97 \times 10^{-7}$ from dispersion, adding at most $10^{-8}$ from the QED effects.
Similar values hold for the 1063 MeV and $5^\circ$ geometry $(S_{\rm exp} =2.8 \times 10^{-7}, \;S_{\rm box} = -2.28\times 10^{-7}, \;S_{\rm coul} = -2.12 \times 10^{-7})$.
The energy distribution of the dispersive spin asymmetry change $S_{\rm box} -S_{\rm coul} \equiv \Delta S_{\rm box}$
for a scattering angle of $5^\circ$ and energies ranging from $650-1150$ MeV is smooth, $|\Delta S_{\rm box}|$ decreasing with energy (see \cite{Jaku24a}).
 Hence our theory is also for $^{208}$Pb at variance with experiment at the GeV energies.

\section{Conclusion}

The radiative corrections to the beam-normal spin asymmetry were estimated with the help of the second-order Born approximation for dispersion and by including the potential for
vacuum polarization and the vertex plus self-energy effect into the Dirac equation, thereby allowing for a nonperturbative treatment of the QED corrections. 
For the two nuclei considered, $^{208}$Pb and $^{12}$C, it is conjectured that the relative QED corrections are  mostly around $2-5\%$ at impact energies $50-200$ MeV and $\vartheta_f \gtrsim 25^\circ$.
However, while $|dS_{\rm QED}|$ increases in the forward hemisphere with energy for $^{12}$C (up to 10\% near 300 MeV),
there exist pronounced diffraction structures for $^{208}$Pb showing no genuine behaviour with energy.

As for dispersion, the present model which considers nuclear excitation energies up to 30 MeV is supposed to give reliable predictions for future experiments in the low-energy regime (50-200 MeV), where hadronic excitations are not yet important.
The dispersive spin asymmetry is  predominantly induced by the high-lying dipole excitations of the target nucleus, particularly for $^{12}$C. However, with increasing angle and collision energy the quadrupole excitations, and for $^{208}$Pb also the octupole excitations, come gradually into play.

A pronounced decrease of the absolute dispersive spin asymmetry change with energy is observed for $^{12}$C at forward angles, particularly between $50-200$ MeV.
For $^{208}$Pb 
$|S_{\rm box} -S_{\rm coul}|$ first decreases, but then increases again beyond some 300 MeV.
Comparing dispersion with vacuum polarization and the vs process, the QED effects constitute for $^{208}$Pb 
 the dominating radiative corrections to the Sherman function, while for $^{12}$C both effects are of  similar magnitude except at small angles where dispersion is most important.

At energies in the GeV region where measurements already are existing, the consideration of the low-lying nuclear excitations plays no role in describing the dispersion effects found in experiment. Also QED effects are  completely negligible at such high energies. Thus the $^{208}$Pb puzzle remains unsolved.

 
\vspace{0.5cm}

\noindent{\large\bf Acknowledgments}

It is a pleasure to thank X.Roca-Maza and M.Gorshteyn for fruitful discussions. I would also like to thank V.Yu.Ponomarev and X.Roca-Maza for calculating the nuclear transition densities for $^{208}$Pb and for  $^{12}$C,  and  H.Spiesberger for forwarding me the ELSEPA code.

\vspace{1cm}

\end{document}